# Electron Identification at CMS detector using LHC data

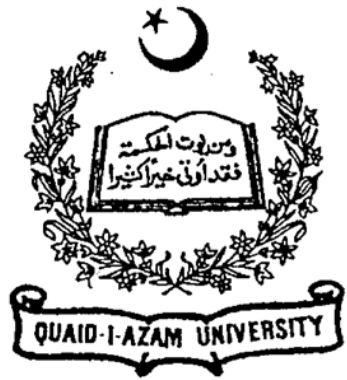

By

# Shoaib Khalid

Department of Physics
Quaid-i-Azam University
Islamabad, Pakistan.
2011

This work is submitted as a dissertation in partial fulfillment for the award of the degree of

MASTER OF PHILOSOPHY
in
PHYSICS

Department of Physics
Quaid-i-Azam University
Islamabad, Pakistan.
2011

# Certificate

It is Certified that the work contained in this dissertation is carried out and completed by under my supervision at department of physics, Quaid-i-Azam University, Islamabad, Pakistan.

<div style="text-align: right;">

Supervised by:

**(Prof. Dr. Hafeez R. Hoorani)**
Director Research
National Center for Physics
Quaid-i-Azam University
Islamabad, Pakistan.

</div>

Submitted through:

**(Prof. Dr. S.K Husnain)**
Chairman
Department of Physics
Quaid-i-Azam University
Islamabad, Pakistan.

*To*

*My family and friends*

# Acknowledgements

I am filled with the praise and glory to All Mighty Allah, the most merciful and benevolent, Who created the universe, with ideas of beauty, symmetry and harmony, with regularity and without any chaos, and gave us the abilities to discover what He thought.

Bless Muhammad (P. B. U. H) the seal of the prophets, and his pure and pious progeny.

First and the foremost, I want to express my deepest gratitude to my venerable teacher and supervisor, Professor Dr. Hafeez R. Hoorani, for his worthy kindness and invaluable favors. He guided my work with infinite patience, enthusiasm and a great amount of tolerance. With all my ventures, I always could rely on his remarkable ability to quickly distill any problem to its essence. His reverential personality and exclusive vision, as a tower of light, guide new researchers not only in Physics but also in other aspects of life.

I am thankful to Chairman, Department of Physics, Quaid-i-Azam University for allowing me to do research in National Center for Physics (NCP).

I would like to thank Prof. Chris Seez (Imperial college London) for supervising me during the two months at CERN and also Nikolaos Rompotis for his guidance and precious time which he gave me during my stay at CERN. My cordial thanks to my friends Ashiq Fareed, Mehar Ali Shah. Their warm companionship make the life at campus beautiful and

*unforgetable. I would like to thank all the members of experimental high energy physics group Dr Irfan Asghar, Dr Shamona Fawad Qazi, Ahmad Abbasi, Muhammad Shoaib, Hamid Ansari, Wajid Ali khan and also Muhammad Athar shah who helped me in one way or the other. I highly value the helping attitude of Taimoor Khurshid who always encouraged me when the load was heavy.*

*I can never thank enough my parents for everything they have done for me. Even then, I wish to record my deepest obligations to my father who always loves me and my mother for her endless contribution to make me stay`on the track. No one tolerates me more than they do. I am forever indebted to the love and caring of my sisters. A great thanks to my family for their unconditional support, belief in me and for keeping me grounded.*

<u>*Shoaib khalid.*</u>

# Abstract

*The Standard Model of particle physics is extremely well tested and yet is not believed to be a theory of everything. Many extensions of the Standard Model predict the existence of new particles. The Large Hadron Collider (LHC) is a high energy proton-proton collider which may be able to produce such particles.*

*In this work strategies have been developed to search for new physics by identifying high energy electrons produced by the decay of W boson using 2.8 pb$^{-1}$ of real data. The electron identification for CMS has been presented. Simple cut based selections and a complete set of variables to distinguish between real electrons and background electrons are described.*

*The detection of electrons is of particular importance at LHC as these particles intervene in the flagship H → ZZ\* → 4e and H → WW\* → 2e2ν channels for the search of the Standard Model (SM) Higgs. Electrons are also important in many SUSY scenarios as produced in the leptonic decays of charginos and neutralinos. Electrons also appears in searches for TeV resonances that may come from new symmetries or as consequences of scenarios involving extra spatial dimensions. Last but not the*

*least, electrons appear in the final state of many Standard Model processes involving top quarks or electroweak bosons, that constitute backgrounds to new signals or are intended to be used as calibration processes.*

*In second part of my thesis background sources for electrons coming from W bosons have been studied. Electron identification variables are plotted in different regions of background, to study the background as deeply as possible. This work has been done for the first time using LHC's real data.*

# Contents

# Chapter 1

# Introduction

The goal of elementary particle physics is to study the smallest constituents of matter and there interaction with each other. To study this fascinating and interesting microcosm, highly complex devices with large dimensions have to be built, firstly to focus energy onto a tiny spot, and then to measure the information from the particles produced, which are generally highly unstable. Paradoxically, the smaller the objects that are studied, the higher is the amount of energy needed, which is why the field is often referred to as high energy physics (HEP). As with all fundamental research, the purpose of elementary particle physics is to seek new knowledge and explore the hidden secrets of this world. This understanding can then inspire applications, for example in nanotechnology, space research, the medical industry, or even novel types of energy production.

## 1.1 Theoretical overview

The experiments over the years have established that smallest constituents are six leptons and six quarks. There interactions are governed by one of the forces. These forces are strong, electromagnetic, and weak. Strong interaction is described by QCD, whereas electroweak model describes electromagnetic and weak interactions. These two theories together form what we call Standard Model.

### 1.1.1 The Standard Model

The Standard Model [1, 2] summarizes the laws of nature and gives predictions with great precision. In the Standard Model, leptons and quarks are classified into three families or generations. We have two quarks in each generation of quarks and similarly two leptons in each generation of leptons. Quarks are fermions having mass and fractional charge. The properties of quarks are such that they can exist, for a short time, only in combinations of two or three quarks known as hadrons. A single quark can not be isolated and also group of four quarks has not been seen yet. The groups of three quarks are called baryons. The groups of two, consisting of a quark and an anti-quark are called mesons. Leptons have one particle like electron and the other a massless and charge less one. Leptons do not have any internal structure. We have an important property of spin which helps us distinguish between particles as fermions or bosons. Fermions which are known as matter particles have intrinsic spin (1/2) while Bosons, the force carriers have intrinsic spin as (0,1). Bosons are responsible for the interactions. Table 1.1 [3] and Table 1.2 [3] show the three generations of quarks and leptons along with their masses and charges respectively.

| Quarks having Intrinsic Spin as 1/2 | | |
|---|---|---|
| Generation | Particle | Charge(e) |
| I | $Up(u)$ | $+2/3$ |
| | $Down(d)$ | $-1/3$ |
| II | $Charm(c)$ | $+2/3$ |
| | $strange(s)$ | $-1/3$ |
| III | $Top(t)$ | $+2/3$ |
| | $Bottom(b)$ | $-1/3$ |

Table 1.1: Standard Model Quarks parameter.

| Leptons having Intrinsic Spin as 1/2 | | | |
|---|---|---|---|
| **Generation** | **Particle** | **Charge(e)** | **Mass (GeV)** |
| I | $Electron(e)$ | $-1$ | $5.1 \times 10^{-4}$ |
|  | $ElectronNeutrino(\nu_e)$ | $0$ | $< 0.8 \times 10^{-8}$ |
| II | $Muon(\mu)$ | $-1$ | $0.105$ |
|  | $MuonNeutrino(\nu_\mu)$ | $0$ | $< 2.7 \times 10^{-4}$ |
| III | $Tau(\tau)$ | $-1$ | $1.777$ |
|  | $TauNeutrino(\nu_\tau)$ | $0$ | $< 0.035$ |

Table 1.2: Standard Model leptons parameter.

In the first generation of quarks, we have Up(u) and Down(d) quarks, in the second generation we have Charm(c) and Strange(s) quarks and in the third, we have Top(t) and Bottom(b) quarks. In the first generation of leptons, we have an Electron(e) and an Electron Neutrino($\nu_e$), in the second generation we have a Muon($\mu$) and a Muon Neutrino($\nu_\mu$) and in the third we have a Tau($\tau$) and a Tau Neutrino($\nu_\tau$). The first generation of fermions ( first generation of quarks and first generation of leptons ) make up our universe as they are the lightest ones. Second and third generation quarks and leptons are heavier and unstable and also decay into lower mass first generation. We can create these heavier particles in high energy experiments. In 1928, Dirac predicted that for every particle there exists an anti-particle of exactly the same mass and spin. If a particle carries an electric charge, then the anti-particle will also carry an equal but opposite charge. Thus, for each quark and each lepton, there is anti-quark and a anti-lepton. An anti-particle is indicated by a bar over its symbol. These particles interact through forces and these forces are governed by the mediation of particles. These mediator particles are known as force carriers and are listed in Table 1.3 [3]. These gauge bosons are the mediators of different types of forces. A particle that can exist in the constraint of uncertainity principle is known as virtual particle and uncertainity principle predicts the range of that particle.

| Gauge Bosons having Spin as 1 | | | |
|---|---|---|---|
| **Particle Name** | **Particle** | **Mass(GeV/c$^2$)** | **Charge (e)** |
| Photon | $\gamma$ | 0 | 0 |
| Gluon | $g$ | 0 | 0 |
| W Boson | $W$ | 80.2 | 1 |
| Z Boson | $Z$ | 91.2 | 0 |

Table 1.3: Standard Model Guage Bosons parameter.

### 1.1.2 Forces in SM

So far the forces of the Standard Model have not been discussed yet. We know that universe is governed by four fundamental forces named as Electromegnetic force, Strong force, Weak force and Gravity but this is not the case in Standard Model. There is no place for gravity in Standard Model yet. Thus the forces which Standard Model includes are Electromagnetic, Weak and Strong force. Below is a description of all these forces.

### 1.1.3 Electromagnetic Force

Electromagnetic force which exists between charges is mediated by photons and with an infinite range as determined by the uncertainity principle. It is represented by U(1) gauge group. Electromagnetic force falls off with the increase in distance as $r^{-2}$. Below is the Feynman diagram for electromagnetic interaction.

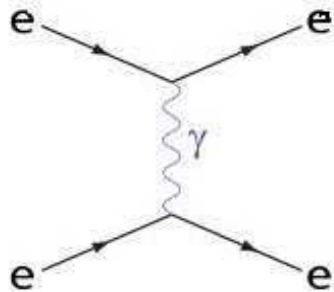

Figure 1-1: Feynman diagram for bhabha scattering

In this figure a electron and a positron annihilate to from a photon which in turn produces a new electron-positron pair.

### 1.1.4 Weak Force

The weak interaction has gauge group SU(2) and it involves the exchange of intermediate vector bosons (W, Z). The range of weak force as predicted in the uncertainity principle is $10^{-18}$m. Weak interaction changes one quark flavor into the other quark flavour. Weak interaction can occur between quarks and leptons because it is independent of color. Leptons having no color do not interact strongly, neutrinos having no charge donot experience electromagnatic force but they can take part in weak interaction. The weak interaction is the only fundamental interaction that does not conserve parity [4, 5] Below are the Feynman diagrams for weak interaction.

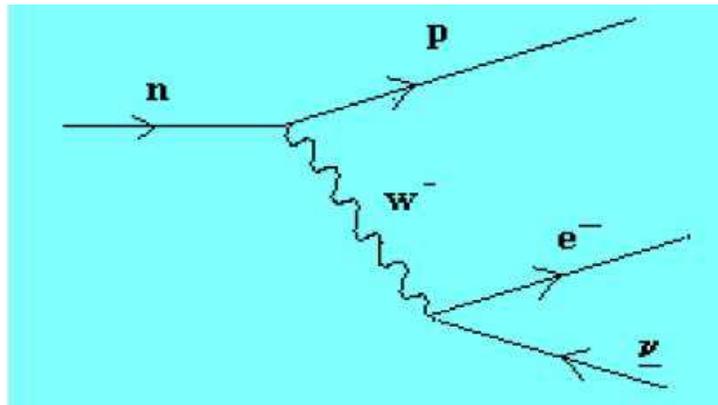

Figure 1-2: Feynman diagram for the beta decay

In the Figure 1.2, a neutron changes to a proton by emitting a W- boson and this W-boson decays to an electron and an electron neutrino. The same beta decay can also be visualized at the quark level and the Figure 1.3 below shows the beta decay at quark level in which one quark changes its flavor and converts into the other.

It is important to note at this point that though gravity is one of the four fundamental forces but it has not been included in the Standard Model yet. There is no theory that unifies gravity with all other forces. Physicists believe that graviton is the mediating particle for this force. Garvity is ignored in all high energy experiments due to its very weak nature.

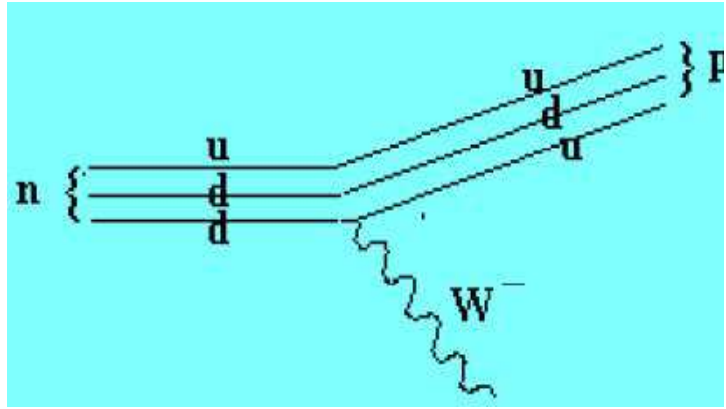

Figure 1-3: Beta decay at the quark level

### 1.1.5 Strong Force

Gluons are mediator of strong force. Gluons interact with those particles which have the property of color charge. Color charge can be red, green and blue for quarks and anti-red, anti-green and anti-blue for anti-quarks. Strong force holds the quarks together to form hadrons. Strong force is represented by the gauge group SU(3) and is the strongest of all the four forces. It does not fall off as the distance increases but infact it increases with the distance. This is the reason why we donot see an isolated quark because by the time the separation is observable, the energy goes far above the pair production energy for quark anti-quark pairs. Quarks have color charge but the composite particles made from quarks do not have color charge. Proton and neutron which constitute the nucleus are made up of quarks. At short range the quarks do not appear to interact very strongly at all. The demonstration of how the asymptotic freedom of the quarks is accommodated in QCD led to the 2004 Nobel Prize in physics [6, 7] .We can visualize strong force by the Feynman diagram given in Figure 1.4. and Figure 1.5. Since gluons carry color, they can interact with each other. In addition to bounded quarks, we can have bounded gluons and is obvious from the Figure 1.6

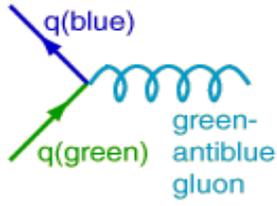

Figure 1-4: Gluon interaction

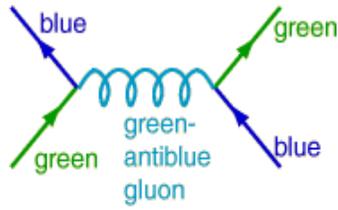

Figure 1-5: Quarks interaction mediated by gluons

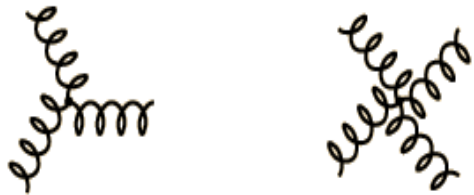

Figure 1-6: Gluons in bound state

## 1.2 The Large Hadron Collider at CERN

### 1.2.1 Design of LHC

The LHC [8] machine has to satisfy stringent design requirements in order to achieve the high luminosity beams required for the physics programmes of the experiments. The maximum available energy for the proton beams is determined by the maximum available super conducting dipole fields necessary to bend the beam in the LHC ring (R=4.5km). These requirements include a proton-proton centre of mass energy of 14TeV and a peak luminosity of L = $10^{34}$cm$^{-2}$s$^{-1}$. Apart from proton-proton operation, the LHC is also colliding heavy-nuclei (Pb-Pb), produced in the existing CERN accelerator complex, giving a centre of mass energy of 1150TeV and peak luminosity of L = $10^{27}$cm$^{-2}$s$^{-1}$.

Since anti-proton would typically require much hour of storage than protons, before injection into the LHC, the use of proton-proton collisions is preferred upon proton anti-proton ones. The LHC is therefore colliding beams of like charged protons, so that, two separate beam-lines are needed to achieve that the two proton beams circulating in opposite direction. The direction of the magnetic field in one beam-line must be opposite to that of the second beam line. This choice however has a disadvantage with respect to the SPS and LEP [9] techniques, where the same set of magnets was used to keep both particles and anti-particles in orbit. CERN decided to solve this problem by using a twin bore magnet, which consists of two sets of coils and beam channels within the same mechanical structure and cryostat. To achieve the LHC energy requirement those dipole magnets(1232 magnets) should be super-conducting (cooled below 2K with super-fluid Helium) with a field strength of approximately 8.33 Tesla. In addition, 392 main quadrapoles are required. Therefore the two proton beams in the LHC machine will have seperate beam pipes which only cross at the interaction points. The performances parameters for p-p operation at the LHC are shown in Table 1.4

| | |
|---|---|
| Circumference | 26.659 km |
| Maximum Dipole field | 8.33 T |
| Magnet Temperature | 1.9 K |
| Beam energy at injection | 450 GeV |
| Beam energy at collision | 7 TeV |
| Nominal Luminosity | $1\times10^{34}$ cm$^{-2}$s$^{-1}$ |
| Number of Bunches | 2808 |
| Number of particles per bunch | $1.15\times10^{11}$ |
| Bunch separation | 24.95 ns |
| Total crossing angle | 285 $\mu$rad |
| Bunch Length (r.m.s.) | 7.55 cm |
| Transverse beam size at Impact Point | 15 $\mu$m $\times$15 $\mu$m |
| Luminosity lifetime | 13.9 h |
| Filling time per ring | 4.3 min |
| Energy loss per turn | 7 KeV |
| Total radiated power per beam | 3.8 kW |
| Stored energy per beam | 362 MJ |

Table 1.4: Parameters of the Large Hadron Collider at CERN, relevant to the pp high luminosity mode.

### 1.2.2 Experiments at LHC

The LHC is a versatile and huge accelerator. It can collide beams of proton with energies around 7-on-7TeV and beam crossing points of unsurpassed brightness, providing the experiments with high interaction rates. The two beams cross each other at four points around the ring, where they can be brought into head-on collision at a centre of mass energy of 14TeV. There are four detectors at these four interaction points. LHC [10] can also collide heavy ion beams such as lead with a total collision energy in excess of 1,250TeV, which is about thirty times higher than at the Relativistic Heavy Ion Collider (RHIC). The research, technical and educational potential of the LHC and its experiments is enormous. In addition to colliding protons, the LHC will smash together ions of lead at speeds which will produce energy densities as high as in the first fraction of a second of the start of the Universe. The LHCb [13] experiment is designed to study the symmetry violations and other rare phenomena in decays of B-meson. This should lead to an understanding of why there is apparently so little antimatter in the Universe, when matter and antimatter should have been produced in equal amounts at the Big Bang. ATLAS [12] and CMS [11] have several functions and they are called general-purpose detectors. In both of these detectors studies are going on for the search of Higgs bosons and Supersymmetric particles. Although their physics goals are quite similar, the design philosophies and the software used in the two detectors are different and so they should provide corroboration of each others results. In ALICE [14] the lead ions are collided to produce a total energy of 1148TeV, where the plasma produced will be studied. A large energy density can be obtained over a wide enough region in the collisions to cause phase transition of nuclear matter into quark-gluon plasma. Studies of such a state of matter are expected to yield important new results. Proton-proton collisions at the LHC will be a copious source of B-mesons. The study of the decay of these mesons will allow a deeper understanding of CP-violation. At a later stage, proton beams from LHC can also be made to collide with electron beams from LEP opening up another field of research. This wide range of physics possibilities will enable LHC to retain its unique place on the frontiers of physics research well into the next century. The complete experimental set-up of LHC and the four detectors is shown in Figure 1.7.

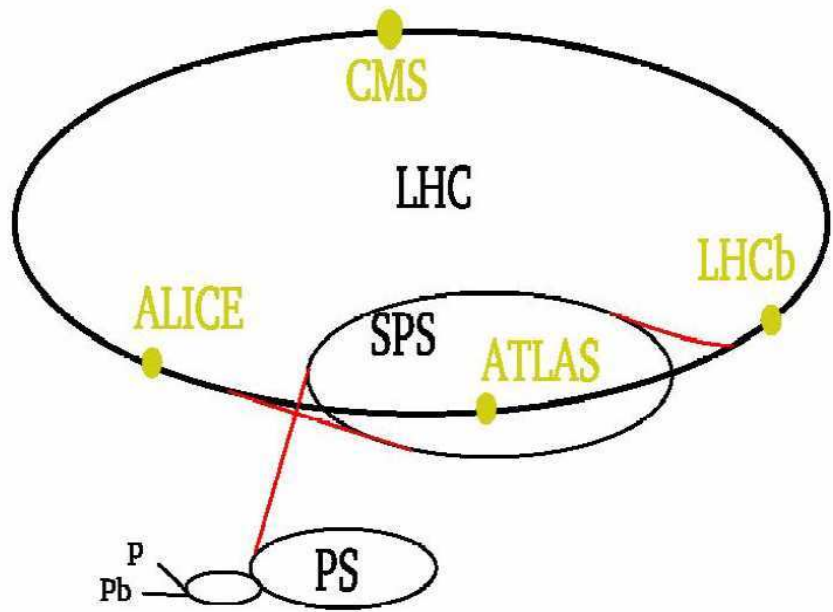

Figure 1-7: Overview of LHC four main detectors.

# Chapter 2

# CMS Detector

CMS stands for Compact Muon Solenoid: compact because it is "small" for its enormous weight, muon for one of the particles it detects, and solenoid for the coil inside its huge superconducting magnet. It is one of the high-energy physics experiment in Cessy, France, part of the Large Hadron Collider (LHC) at CERN. The CMS detector [15] is designed to see a wide range of particles and phenomena produced in high-energy collisions in the LHC. It is like a cylindrical onion, different layers of detector stop and measure the different particles, and use this key data to build up a picture of events at the heart of the collision.

Using this data scientists are searching for new phenomena that will help to answer questions such as: What is the Universe really made of and what forces act within it? And what gives everything substance? CMS will also measure the properties of previously discovered particles with unprecedented precision, and be on the lookout for completely new, unpredicted phenomena.

## 2.1 Design of CMS

Detectors consist of layers of material that exploit the different properties of particles to catch and measure the energy and momentum of each one. CMS was designed around getting the best possible scientific results [16], and therefore to look for the most efficient ways of finding evidence for new physical theories. This put certain requirements on the design. CMS needed:

- a system with high performance for detecting and measuring muons.

- a high resolution system to detect and measure electrons and photons (an electromagnetic calorimeter).

- a high quality central tracking system to give accurate momentum measurements.

- a "hermetic" hadron calorimeter, designed to entirely surround the collision and prevent particles from escaping.

With these priorities in mind, the important thing was a very strong magnet. The higher a charged particle's momentum, the less its path is curved in the magnetic field, so when we know its path we can measure its momentum. A strong magnet was therefore needed to allow us to accurately measure even the very high momentum particles, such as muons. A large magnet also allowed for a number of layers of muon detectors within the magnetic field, so momentum could be measured both inside the coil (by the tracking devices) and outside of the coil (by the muon chambers).

The magnet is the "Solenoid" in Compact Muon Solenoid (CMS). The solenoid is a coil of superconducting wire that creates a magnetic field when electricity flows through it; in CMS the solenoid has an overall length of 13m and a diameter of 7m, and a magnetic field about 100,000 times stronger than that of the Earth. It is the largest magnet of its type ever constructed and allows the tracker and calorimeter detectors to be placed inside the coil, resulting in a detector that is, overall, "compact", compared to detectors of similar weight.

The design of the whole detector was also inspired by lessons learnt from previous CERN experiments at LEP (the Large Electron Positron Collider). Engineers found that building sections above ground, rather than constructing them in the cavern with all its access and safety issues, saved valuable time. Another important conclusion was that sub-detectors should be made more easily accessible to allow for easier and faster maintenance.

Thus CMS was designed in fifteen separate sections or "slices" that were built on the surface and lowered down ready-made into the cavern. Being able to work in parallel on excavating the cavern and building the detector saved valuable time. This slicing, along with the careful design of cabling and piping, also ensures that the sections can be fully opened and closed with minimum disruption, and each piece remains accessible within the cavern.The sliced view of CMS is shown in Figure 2.1.

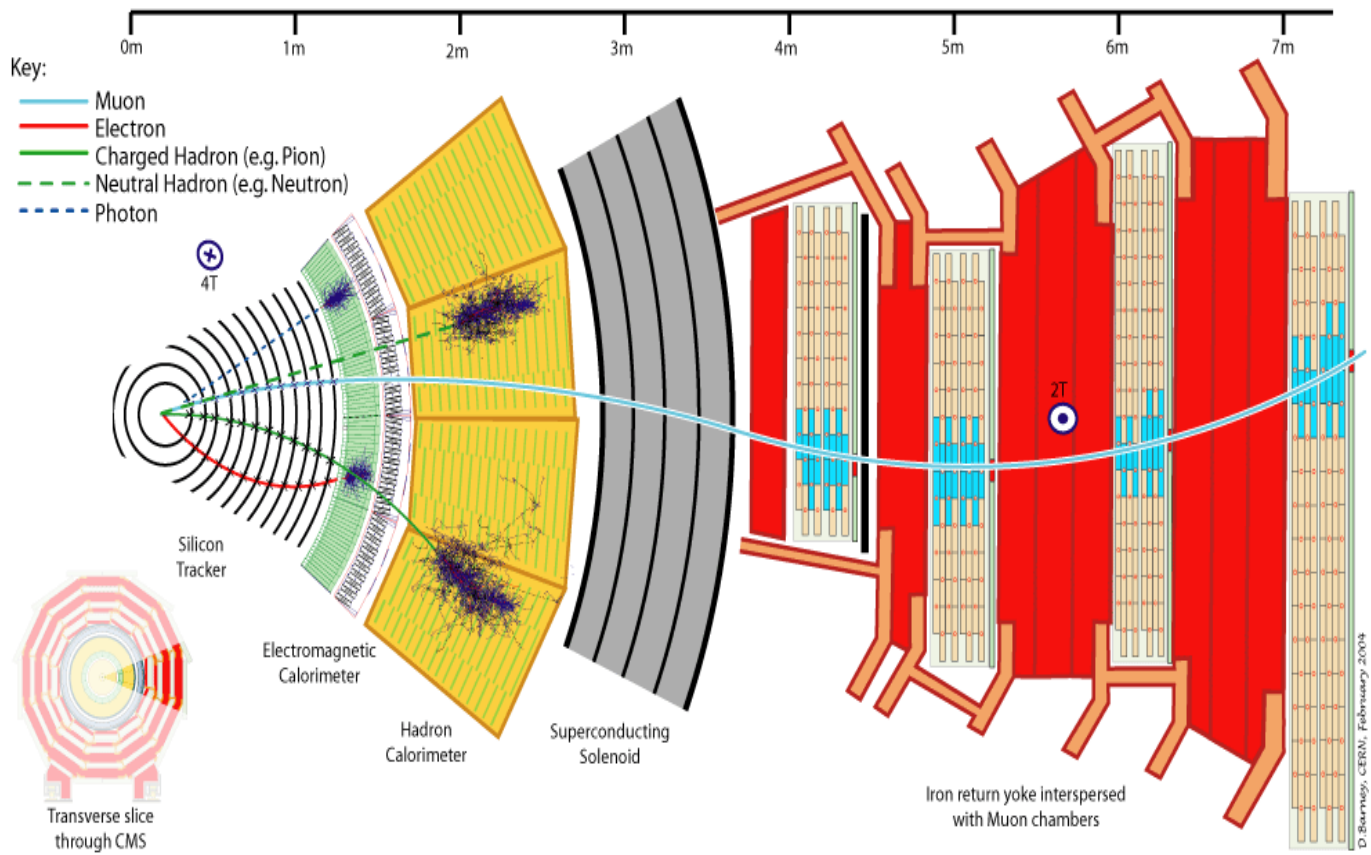

Figure 2-1: Transverse Slice of the Compact Muon Solenoid (CMS) Detector

### 2.1.1 Overview of CMS detector

The CMS experiment is 15m wide, 21m long and 15m high, and sits in a cavern that could contain all the residents of Geneva; albeit not comfortably. The detector is like a giant filter, and its each layer is designed to stop, track or measure a different type of particle emerging from proton-proton and heavy ion collisions. Finding the momentum and energy of a particle gives clues to its identity, and particular patterns of particles or "signatures" are indications of new and exciting physics. The detector is built around a huge solenoid magnet. This takes the form of a cylindrical coil of superconducting cable, cooled to $-268.5^oC$, that generates a magnetic field of 4 Tesla, which is 100,000 times that of the Earth. Particles that emerges from collisions first meet a tracker, made entirely of silicon, that charts their positions as they move through the detector, allowing us to measure their momentum. Outside the tracker there are calorimeters that measure the energy of particles. In measuring the momentum, the tracker should interfere with the particles as little as possible, whereas the calorimeters are specifically designed to stop the particles in their tracks.

The Electromagnetic Calorimeter (ECAL) is made of lead tungstate, a very dense material that produces light when photons and electrons passess through it, and measures the energy of photons and electrons whereas the Hadron Calorimeter (HCAL) is designed principally to detect any particle made up of quarks (the basic building blocks of protons and neutrons). The size of the magnet allows the tracker and calorimeters to be placed inside its coil, resulting in an overall compact detector.

As the name suggests CMS is also designed to measure muons. The outer part of the detector, the iron magnet "return yoke", which guides and confines the magnetic field, also stops all remaining particles but muons and weakly interacting particles, such as neutrinos, from reaching the muon detectors. Four layers of detectors are interleaved with the iron, which also provides the detector's support structure. Within the LHC, bunches of particles collide up to 40 million times in a second, so a "trigger" system is introduced that saves only potentially interesting events. This reduces the number recorded from one billion to around 100 per second. The general layout of CMS detector is shown in Figure 2.2.

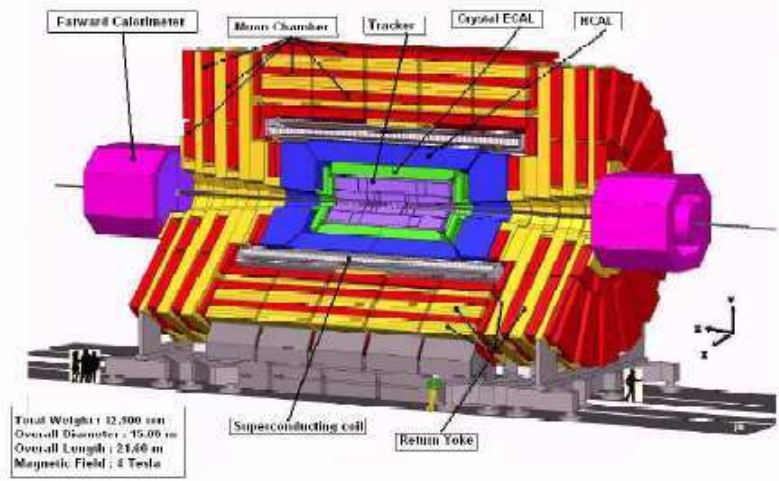

Figure 2-2: General layout of the CMS detector at the Large Hadron Collider at CERN.

## 2.2 Geometry of CMS

### 2.2.1 The Magnet

The CMS magnet is the central device around which the experiment is built, with a magnetic field of 4 Tesla and 100,000 times stronger than the Earth's. Its job is to bend the paths of particles coming from high-energy collisions in the LHC. The more momentum a particle has the less its path is curved by the magnetic field, so tracing its path gives a measure of momentum. CMS began with the aim of having the strongest magnet possible because a higher strength field bends paths more and combined with high-precision position measurements in the tracker and muon detectors, this allows precise measurement of the momentum of even high-energy particles.

The CMS magnet is a "solenoid". A solenoid is a magnet made of coils of wire which produce a uniform magnetic field when electricity flows through them. The CMS magnet is "superconducting", which allow electricity to flow without resistance and creating a strong magnetic field. In fact at ordinary temperatures the strongest possible magnet has only half the strength of the CMS solenoid [17].

The tracker and calorimeter detectors (ECAL, HCAL) fit snugly inside the magnet coil while the muon detectors are interleaved with a 12-sided iron structure that surrounds the magnetic coils and contains and guides the field. Made up of three layers this "return yoke" reaches out 14 metres in diameter and also acts as a filter, which allows through only muons and other weakly interacting particles such as neutrinos. This enormous magnet also provides most of the experiment's structural support, and will be very strong itself to withstand the forces of its own magnetic field.

### 2.2.2 The Tracker

Momentum of particles is important in helping us to build up a picture of events at the heart of the collision. One way to calculate the momentum of a particle is to track its path through a magnetic field; the more curved the path, the less momentum the particle had. The CMS tracker records the paths taken by charged particles by finding their positions at number of key points. The tracker [18] can reconstruct the paths of high-energy muons, hadrons (particles made up of quarks) and electrons as well as tracks coming from the decay of very short-lived particles such as beauty or "b quarks" that will be used to study the differences between matter and antimatter.

The tracker needs to record particle paths accurately and in the mean time should be lightweight so as to disturb the particle as little as possible. It does this by taking position measurements so accurate that tracks can be reliably reconstructed just using a few measurement points. Each measurement is accurate to about 10 $\mu$m, a fraction of the width of a human hair. It is also the inner most layer of the detector and so receives the large volume of particles: the construction materials were therefore carefully chosen to resist radiation.

The final design consists of a tracker made completely of silicon: the pixels, at the very core of the detector and dealing with the highest intensity of particles, and the silicon microstrip detectors that surround it. As particles travel through the tracker the pixels and microstrips produce tiny electric signals that are amplified and then detected. The tracker employs sensors covering an area the size of a tennis court, with 75 million separate electronic read-out channels: in the pixel detector there are about 6000 connections per square centimetre.

### 2.2.3 The Electromagnetic Calorimeter (ECAL)

To build up a picture of events that are occurring in the LHC, CMS must be able to find the energies of emerging particles. Of particular interest are electrons and photons, because of their use in finding the Higgs boson and other new physics also. Electrons and photons are measured using an electromagnetic calorimeter (ECAL) [19]. But to find them with the necessary precision in the very strict conditions of the LHC, a high magnetic field, high levels of radiation and only 25 nanoseconds between collisions required very particular detector materials. Lead tungstate crystal is made primarily of metal and is heavier than stainless steel, but with a touch of oxygen in this crystalline form it is highly transparent and "scintillates" when photons and electrons pass through it. It produces light in proportion to the particle's energy. These high-density crystals produces light in fast, short, well-defined photon bursts that allow for a precise, fast and fairly compact detector.

Photodetectors that have been especially designed to work within the high magnetic field,

are also glued onto the back of each of the crystals to detect the scintillation light and convert it to an electrical signal that is amplified and than sent for analysis. The ECAL, made up of a barrel section and two "endcaps", forms a layer between the tracker and HCAL. The cylindrical "barrel" consists of 61,200 crystals formed into 36 "supermodules", each weighing around three tonnes and also containing 1700 crystals. The flat ECAL endcaps seal off the barrel at either end and are made up of about 15,000 further crystals. For extra spatial accuracy, the ECAL also contains Preshower detectors that sit in front of the endcaps. These allow CMS detector to distinguish between single high-energy photons (often signs of exciting physics) and the less interesting close pairs of low-energy photons.

### 2.2.4 The Hadron Calorimeter

The Hadron Calorimeter (HCAL) measures the energy of "hadron", particles made of quarks and gluons (for example protons, neutrons, kaons and pions). Additionally it also provides indirect measurement of the presence of non-interacting, uncharged particles such as neutrinos. Measuring these particles is important as they can tell us if new particles such as the Higgs boson or supersymmetric particles (much heavier versions of the standard particles we know) have been formed. As these particles decay they may produce new particles that do not leave any record of their presence in any part of the CMS detector. To spot these particles the HCAL must be "hermetic", that is make sure it captures, to the extent possible, every particle emerging from the collisions. This way if we see particles shoot out one side of the detector, but not the other, with an imbalance in the momentum and energy (measured in the sideways "transverse" direction relative to the beam line), we can deduce that we're producing some "invisible" particles. To ensure that we are seeing something new, rather than just letting familiar particles escape undetected, layers of the HCAL were built in a staggered fashion so that there are no gaps in direct lines that a familiar particle might escape through.

The HCAL is a sampling calorimeter, meaning it finds a particle's position, energy and arrival time using alternating layers of "absorber" and fluorescent "scintillator" materials that produce a rapid light pulse when the particle passes through it. Special optic fibres collect up this light and feed it into readout boxes where photodetectors are used to amplify the signal. When the amount of light in a given region is added up over many layers of tiles in depth, called a "tower", this total amount of light is a measure of a particle's energy. As the HCAL is thick and massive, fitting it into "compact" CMS was a challenge, as the cascades of particles produced when a hadron hits the dense absorber material (known as showers) are large, and the minimum amount of material needed to contain and measure them is about one metre.

To accomplish this feat, the HCAL is organised into barrel (HO and HB), endcap (HE) and forward (HF) sections. There are 36 barrel "wedges", each weighing 26 tonnes. These form the last layer of detector inside the magnet coil while a few additional layers, the outer barrel (HO), sit outside the coil, ensuring no energy leaks out the back of the HB undetected. Similarly, 36 endcap wedges measure particle energies as they come through the ends of the solenoid magnet. Lastly, the two hadronic forward calorimeters (HF) are positioned at either end of CMS, to pick up the large number of particles coming out of the collision region at shallow angles relative to the beam line. These receive the bulk of the particle energy contained in the collision so must be very resistant to radiation and use different materials to the other parts of the HCAL [20].

### 2.2.5 The Muon System

As the name "Compact Muon Solenoid" suggests, detecting muons is one of CMS's most important task. Muons are charged particles that are just like electrons and positrons, but are 200 times heavier than electrons. We expect them to be produced in the decay of a number of potential new particles; for example, one of the clearest "signatures" of the Higgs Boson is its decay into four muons. Because the muons can penetrate several metres of iron without interacting, unlike most particles they are not stopped by any of CMS's calorimeters. Therefore, chambers to detect muons are placed at the very edge of the experiment where they are the only particles that are likely to register a signal. A particle is measured by fitting a curve to hits among the four muon stations [21], which sit outside the magnet coil and are interleaved with iron "return yoke" plates. By tracking its position through the multiple layers of each station, combined with tracker measurements the detectors precisely trace a muon's path. This gives a measurement of its momentum because we know that particles travelling with large momentum bend less in a magnetic field. As a consequence, the CMS magnet is powerful enough so we can bend even the paths of very high-energy muons and calculate their momenta.

In total there are 1400 muon chambers: 250 drift tubes (DTs) and 540 cathode strip chambers (CSCs) track the particle's positions and provide a trigger, while 610 resistive plate chambers (RPCs) form a redundant trigger system, which decides quickly to keep the acquired muon data or not. Because of the many layers of detector and different specialities of each type, the system is naturally robust and able to filter out the background noise. DTs and RPCs are arranged in concentric cylinders around the beam line ("the barrel region") while CSCs and RPCs, make up the "endcaps" disks that cover the ends of the barrel.

## 2.3 The Trigger system

While pp collisions occur at a rate of 40MHz, it is almost impossible to store on tape all the information about every collision, practical and technical difficulties impose a limit of about 100Hz in the acceptable rate of permanently stored data. Furthermore, the rate of interesting events is smaller by order of magnitudes than the total interaction rate, hence a trigger system is built up with the two fold task to reject a factor $4 \times 10^5$ of the collisions and to select in a short time the interesting physics events with big efficiency. The CMS has two-step trigger and data acquisition system, which is designed to inspect the detector output at full bunch crossing frequency and to store the selected events at the maximal allowed storage rate of 100Hz. The first level trigger (L1) is a hardware trigger designed to reduce the rate of accepted events to less than 100KHz. It is based on the identification of muons, electrons, photons, jets and missing transverse energy. The second level trigger or HLT (High Level Trigger) is a sofware trigger, which is running on the dedicated farm of commercial processors and designed to reduce the maximum L1 output rate to the final output rate of 100Hz. The bandwidth of 100Hz should then be shared between all the channels of interest.

### 2.3.1 Level-1 Trigger

The Level-1 trigger selection is based exclusively on the information of calorimeter and muon chamber, processed with hardware logical circuits, though with coarse granularity. The Level-1 trigger [22] system is required to be capable of processing every 40MHz pp collision and reduce to 100kHz the data rate to pass to the HLT. At LHC startup the CMS Level-1 output rate will be reduced to 50kHz for low luminosity and it will be raised to the designed 100kHz at full LHC luminosity.

### 2.3.2 High-Level Trigger

The High-Level trigger selection (HLT) is realized with a software running on a farm of commercial processors. The goal of HLT [23] is to reduce the Level-1 output rate to 100Hz mass storage with dedicated "fast" algorithms.

## 2.4 The Grid

Even when whittled down by the trigger system, CMS still produces a large amount of data that must be analysed, more than five petabytes per year when running at peak performance. To meet this challenge, the LHC employs a novel computing system, a distributed computing and data storage infrastructure which is called the Worldwide LHC Computing Grid (WLCG) [24]. In 'The Grid', tens of thousands of standard PCs collaborate worldwide to have much more processing capacity than could be achieved by a single supercomputer, giving access to data to thousands of scientists all around the world. The "Tier 0" centre at CERN first reconstructs the full collision events and analysts start to look for patterns; but the data has a long way to go yet.

Once CERN has made a primary backup of the data it is then sent to large "Tier 1" computer centres in seven locations around the world: in France, Spain, Italy, Germany, Taiwan, the UK and the US. Here events are reconstructed again, using information from the experiment to improve calculations using refined calibration constants. Tier 1 starts to interpret and make sense of the particle events and collate the results to see emerging patterns. Meanwhile each sends the most complex events to a number of "Tier 2" facilities, which total around 40, for further specific analysis tasks. In this way information branches out from each tier across the world so that, on a local level, students and physicists whether in Rio de Janeiro or Oxford, can study CMS data from their own computer, updated on a regular basis by the LHC Computing Grid.

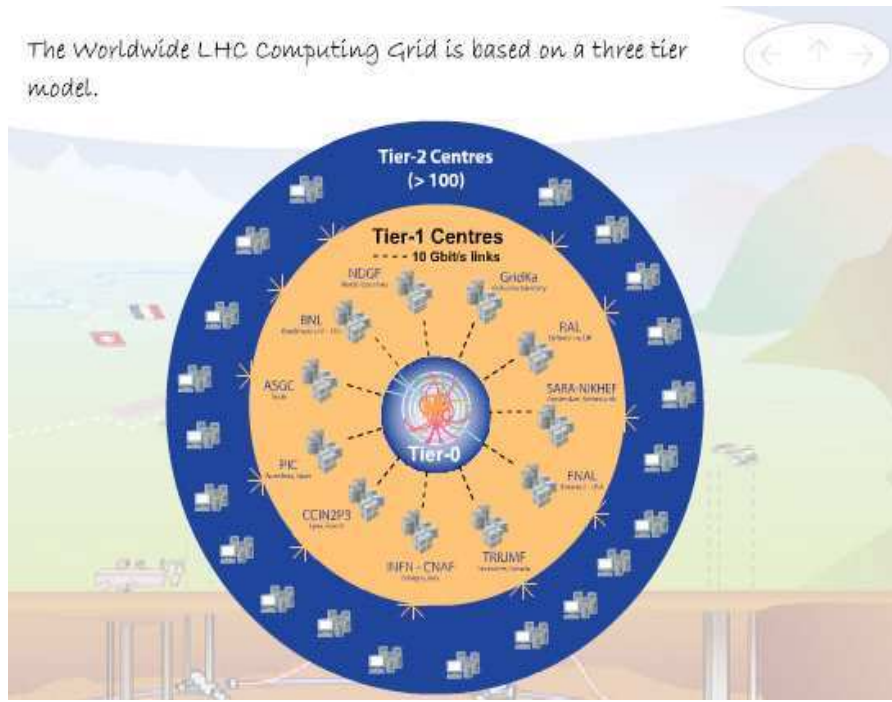

Figure 2-3: The LHC tiered computing model to distribute data around the world

## 2.5 Software Package

### 2.5.1 Root

After many years of experience in developing interactive data analysis systems like PAW [25], it was realized that the growth and maintainability of these products, written in FORTRAN and using 20-year-old libraries, had reached its limits. Although still popular in the physics community, these systems do not match up to the challenges offered by the next generation particle accelerator, the Large Hadron Collider (LHC) at CERN, in Geneva, Switzerland. The expected amount of data that will be produced by the LHC is of the order of several petabytes (1PB = 1,000,000GB) per year. This is about two to three orders of magnitude more than what is being produced by the current generation of accelerators. This was the motivation to build ROOT framework. ROOT is an object-oriented framework aimed at solving the data analysis challenges of highenergy physics. It was originally designed for particle physics data analysis and contains several features specific to this field, but it is also commonly used in other applications where large amounts of data need to be processed such as data mining and astronomy. It is written in C++ language. The ROOT package [26] contains many useful functions and tools that make our analysis comprehensive and easy. Some of the important and useful functions are histogramming and graphing to visualize and analyze distributions and functions, curve fitting and minimization of functionals, statistics tools used for data analysis, matrix algebra, four-vector computations, as used in high energy physics, standard mathematical functions, 3D visualizations (geometry) and interfacing Monte Carlo event generators. A key feature of ROOT is a data container called tree, with its substructures branches and leaves. ROOT's focus on performance is caused by the amount of data that the Large Hadron Collider's experiments will collect, estimated to several petabytes per year. Physicists are expected to analyze this data using ROOT. ROOT is currently mainly used in data analysis and data acquisition in high energy physics experiments most current experimental plots and results are obtained using ROOT. Many particle accelarators use ROOT for the physics analysis e.g. BaBar, CDF [27], PHENIX and also detectors as ALICE, ATLAS, CMS, LHCb.The inclusion of the CINT C++ interpreter makes this package very versatile as it can be used in interactive, scripted and compiled modes in a manner similar to commercial products like MATLAB [28].

## Chapter 3

# Particle Identification

Particle identification is the process of using information left by a particle passing through a particle detector to identify the type of particle. Particle identification reduces backgrounds and improves measurement resolutions, and is essential to many analyses at particle detectors. The defining characteristics of any particle are its mass and its charge. The charge of a particle is easily determined by its behavior in a magnetic field. The magnets create a magnetic field around the collision-event that sweeps positive and negative particles in opposite directions, allowing us to determine the charge.

## 3.1 Charged particle

Charged particles have been identified using a variety of techniques. All methods rely on a measurement of the momentum in a tracking chamber combined with a measurement of the velocity to determine the charged particle mass, and therefore its identity.

### 3.1.1 Ionization loss of charged particles

The detection of nuclear particles depends ultimately on the fact that, directly or indirectly, they transfer energy to the medium they are traversing via the process of ionization or excitation of the constituent atoms. The Bethe-Bloch formula [29] for the mean rate of ionization loss of a charged particle is given by

$$\frac{dE}{dx} = \frac{4\pi N_o z^2 e^4}{mv^2} \frac{Z}{A} \left[ \ln\left(\frac{2mv^2}{I(1-\beta^2)}\right) - \beta^2 \right]$$

where

- $m$ is the electron mass,
- $z$ and $v$ are the charge (in units of e) and velocity of the particle,
- $\beta = v/c$,
- $N_o$ is Avogadro's number,
- $Z$ and $A$ are the atomic number and mass of the atoms in the medium,
- $x$ is the path length in the medium (g cm$^{-2}$),
- $I$ ( $\sim 10\ ZeV$) is the effective ionization potential averaged over all electrons.

The bulk of the energy loss results from the formation of ion pairs (a positively charged ion and an electron) in the medium. The electrons knocked out by the incident particle may themselves produce secondary ionization . The total number of ions pairs is proportional to the energy loss.

### 3.1.2 Radiation loss of charged particles

Charged particles lose energy in traversing a medium in two ways, the ionization energy loss which we discussed earliar and the process of radiation loss or bremsstrahlung. The radiative energy loss of the particles occur principally with the atomic nuclei of the medium. The nuclear electric field decelerates the particle, and the energy change appears in the form of a photon. The energy loss due to bremsstrahlung radiation for an electron traversing thickness dx of a medium is given by,

$$\left[\frac{dE}{dx}\right]_{rad} = \frac{-E}{X_o}$$

where $X_o$ , the radiation length, is given by:

$$X_0 = \frac{716.4A}{Z(Z+1)\ln\left(287/\sqrt{Z}\right)}$$

The radiation length $X_o$ may simply be defined as that thickness of the medium which reduces the mean energy of a beam by $1/e$ (2.718). For example, the radiation lengths in lead and lead tungstate ($PbWO_4$) are 0.56cm and 0.89cm respectively.

Energy loss by ionization is dominant at low energy. For relativistic particles the energy loss by ionization varies slowly (logarithmically) with energy. At sufficiently high energies radiative losses become more important for all charged particles. Electrons lose energy by bremsstrahlung at a rate approximately proportional to their energy. The energy at which which the loss rates from ionization and bremsstrahlung are equal is defined as the 'critical energy' ($E_c$). For electrons this energy is given by:

$$E_c \approx \frac{600 MeV}{Z}$$

## 3.2 Kinematics variables used

Some of the variables which are used in electron identification [30] are defined below:

- $\eta = -\log \tan\left(\frac{\theta}{2}\right)$
- SuperCluster energy over track momentum at vertex $(E/P)$
- DeltaEta between SuperCluster position and track direction at vertex extrapolated to Ecal($\Delta\eta$).
- DeltaPhi between SuperCluster position and track direction at vertex extrapolated to Ecal($\Delta\phi$).
- Ratio of energy in HCAL behind SuperCluster to SuperCluster energy $(H/E)$
- Bremstralung fraction ={(track momentum at vertex - track momentum at ECAL)/(track momentum at vertex)}$(f_{brem})$
- RMS shower width variable in $\eta$ direction $(\sigma_{i\eta i\eta})$
- No of allowed missing hits between the vertex and the first track hit.($missinghits$)
- Difference in dipole angle between electron track and accompanying track($\cot\theta_{electron} - \cot\theta_{accompanyingtrack}$), $(D\cot\theta)$
- Distance of closest approach of the accompanying tracks to the electron track $(Dist)$.
- Sum of all the tracks pt over electron Et in a cone with $\Delta R < 0.3$ centred at the track vertex, $(\sum Pt/Et)$, $(trackisolation)$.

• Sum of all the tracks Et over electron Et in a cone with $\Delta R < 0.3$ centred at the supercluster postion in the Ecal, $(\sum Et/Et), (Ecal isolation)$.

• Sum of all the tracks Et over electron Et in a cone with $\Delta R < 0.3$ centred at the supercluster postion in the Hcal, $(\sum Et/Et), (Hcal isolation)$

## 3.3 Algorithms used for electron Identification

Cut based electron identification [31] method has been used to seperate the real electrons, but before coming to this an introduction of a few more methods has been given which can also be used as an electron identification tool in the future. Some of these techniques are listed below:

### 3.3.1 Electron identification using likelihood

The following variables are currently used to discriminate between real and fake electrons in the likelihood [32, 33] :

• Energy of closest BasicCluster to track impact point at Ecal / outermost track momentum.

• DeltaEta between SuperCluster position and track direction at vertex extrapolated to Ecal$(\Delta \eta)$.

• DeltaPhi between SuperCluster position and track direction at vertex$(\Delta \phi)$

• Ratio of energy in HCAL behind SuperCluster to SuperCluster energy $(H/E)$

• Cluster shape variables:

· Energy in 3x3 crystals / energy in 5x5 crystals

· RMS shower width variable in $\eta$ direction $(\sigma_{i\eta i\eta})$

### 3.3.2 Electron identification using Neural Networks

The electron identification with Neural Network (NN) [34] is a standard CMS tool for electron identification along with cut-based electron identification and likelihood-based electron identification. To limit the complexity of the network, it uses only a minimal set of variables to perform the electron identification. The following variables are currently used to discriminate between real and fake electrons in the NN:

• (1/E - 1/P)

- Ratio of energy in HCAL behind SuperCluster to SuperCluster energy ($H/E$)
- DeltaPhi between SuperCluster position and track direction at vertex ($\Delta\phi$)
- RMS shower width variable in $\eta$ direction ($\sigma_{i\eta i\eta}$)

### 3.3.3 Cut based electron identification

Electrons from W decay, can be distinguished from other particles due to their unique characteristics that are primarily measured in the ECAL and tracker. Ideally, the electron track would match well with the cluster of energy found in the ECAL, both in position and momentum [35]. The track would also emanate directly from the event vertex and be isolated from other tracks and calorimeter energy deposits. Electron identification makes use of a complete set of variables to distinguish between real electrons and background electrons, One thing important to be mentioned here is that we are taking electrons coming from W boson($w-> e\nu$) as real electrons and all the electrons coming from other sources will be considered as background, one reason could be that the real data(2.8 pb$^{-1}$) which is used in this thesis has a larger W cross section, and also the W channel contains large number of high pt electron. To achieve a high efficiency and purity for electron candidates and high exclusion of background, the electron identification criteria have been designed and mainly divided into 3 types of categories [30]:

**1) Electron identification variables**  The electron identification variables includes spatial matching between track and ECAL supercluster $\Delta\phi$ and $\Delta\eta$, and Calorimeter shower shape variables [36] : the width of the ECAL cluster along the $\eta$ direction computed for all the crystals in the 5×5 block of crystals centered on the highest energy crystal of the seed cluster $\sigma_{i\eta i\eta}$. These variables are helpful in identifying the electron, as the electron track must match with the supercluster, So ideally $\Delta\phi$ and $\Delta\eta$ should be zero for a electron. $\Delta\phi$ and $\Delta\eta$ are large for those particles whose track and ECAL supercluster badly match with each other, this might be due to the radiation which they lose along there path.

**2) Electron isolation variables**  The performance of the electron identification depends of course on the degree of isolation imposed on the electron candidates. The variables which are used to isolate the electron tracks from other tracks are $H/E = E_{Hcal}/E_{Ecal}$, It is the ratio of the energy deposits ( $E_{Hcal}$ and $E_{Ecal}$) in the HCAL and ECAL calorimeters respectively.

$E_{Hcal}$ being computed from the energy deposits in the HCAL within a cone of radius $\Delta R = \sqrt{\Delta\phi^2 + \Delta\eta^2}$<0.15 around the ECAL supercluster direction. The ratio of hadronic to electromagnetic energy of an electron is a very good criteria to discriminate electron and jet, since at low energies, electrons will be fully contained in the electromagnetic calorimeter (ECAL) and deposit very little energy in the hadronic calorimeter (HCAL), while at high energies electrons will deposit most of the energy in the ECAL and deposit some energy in the HCAL. On the other side the jet will leak more energy in the Hcal due to longer maximum depth of hadronic showers. The variables ECAL isolation and HCAL isolation which are explained earlier are also use to isolate the electron track.

**3) Coversion rejection variables**  Due to the non-negligible material budget of the CMS tracking system, large multiple scattering, bremsstrahlung and high photon conversion rates are all prevalent. Electrons from photon conversions constitute approximately 15-35% of electrons in QCD events, depending on the cuts applied. Electrons from photon conversions are therefore a non-negligible background to prompt electrons from hadron collisions and must be rejected efficiently. To that end, we have developed three cuts designed to reject such electrons.

• Search for the Conversion Partner Track: the tracks of the resulting electrons from a conversion decay are parallel to each other at the decay point, and remain so in the rz plane. This is a unique feature that is the basis of the algorithm we use. To exploit this geometry, all Combinatorial Track Fitter (CTF) tracks within a cone of $\Delta R < 0.3$ around the electron GSF track and with charge opposite that of the GSF track, are pre-selected. For each of these tracks, the following two quantities are defined:

– $D\cot\theta = \cot\theta_{CTFtrack}$ - $\cot\theta_{GSFtrack}$

– The Dist is defined as the two-dimensional distance (xy plane) between the two tracks when the CTF track in question and the electron GSF track would be parallel when extrapolated. We donot know the point where a photon converts to a $e^- e^+$. So we take a nearest accompanying track, if the distance between the electron track and accompanying track is below certain value then we say that this electron is coming from conversion. This distance is calculated analytically by a simple intersection of helices method using the track parameters of the two tracks as input. Figure 3.1 shows the definition of dist, as well as the sign convention used. Requiring that the $|cot(\theta)| < 0.02$ and $|Dist| < 0.02$cm [37] efficiently rejects a significant portion of the remaining electrons from photon conversions.

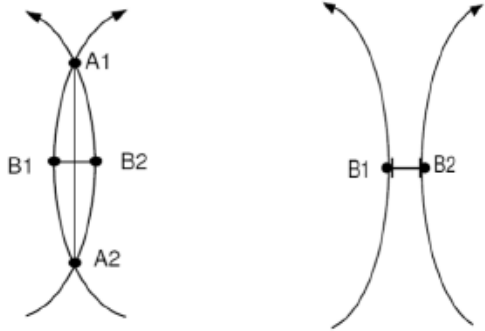

Figure 3-1: The Dist quantity is the two dimensional distance between points B1 and B2 in the xy plane as seen above. At these points, the two tracks from the photon conversion are parallel. The Dist is defined to be negative when the two tracks overlap, and is positive otherwise.

- Missing hits : Photon conversions occur later inside the tracker volume and not at the primary vertex. Therefore, the first valid hit of a resulting electron track may not necessarily be located in the innermost tracker layer. We call a hit valid if it is used in the final out of the track. Extrapolating the track of an electron from a photon conversion back to the beam-line, one could cross active detector layers which do not have hits compatible with the track (in other words, a missing hit). For prompt electrons, whose trajectories start from the beam-line, we do not expect any missing hits in the crossed inner tracker layers. We can therefore use this expectation of no missing hits at inner radii to reject electrons from photon conversions. Determining whether a track has missing hits is possible via its associatied Hit Pattern object. We find that requiring the number of expected layers with a missing hit be $\leq 1$ efficiently rejects a large fraction of electrons from photon conversions.

### 3.3.4 Cut values for cut based method

Cuts are applied on all of these ten variables in both barrel and endcap regions seperately to minimize the background with smallest loss of signal efficiency. The barrel region corresponds to $\Delta\eta < 1.479$, and endcap region corresponds to $\Delta\eta > 1.479$. A set of cut on all of these ten variables corresponds to a working point. Different working points are shown in the Table 3.1 [31] .

| $Efficiencies$ | 95% | 90% | 85% | 80% | 70% | 60% |
|---|---|---|---|---|---|---|
| $missing hits \leq$ | 1 | 1 | 1 | 0 | 0 | 0 |
| $Dist$ | N/A | 0.02 | 0.02 | 0.02 | 0.02 | 0.02 |
| $D \cot \theta$ | N/A | 0.02 | 0.02 | 0.02 | 0.02 | 0.02 |
| **BARREL** | | | | | | |
| $Relative isolation$ | | | | | | |
| $Track isolation$ | 0.15 | 0.12 | 0.09 | 0.09 | 0.05 | 0.04 |
| $Ecal isolation$ | 2.00 | 0.09 | 0.08 | 0.07 | 0.06 | 0.04 |
| $Hcal isolation$ | 0.12 | 0.10 | 0.10 | 0.10 | 0.03 | 0.03 |
| $Electron ID$ | | | | | | |
| $\sigma_{i\eta i\eta}$ | 0.01 | 0.01 | 0.01 | 0.01 | 0.01 | 0.01 |
| $\Delta\phi$ | 0.8 | 0.8 | 0.06 | 0.06 | 0.03 | 0.025 |
| $\Delta\eta$ | 0.007 | 0.007 | 0.006 | 0.004 | 0.004 | 0.004 |
| $H/E$ | 0.15 | 0.12 | 0.04 | 0.04 | 0.025 | 0.025 |
| **ENDCAPS** | | | | | | |
| $Relative isolation$ | | | | | | |
| $Track isolation$ | 0.08 | 0.05 | 0.05 | 0.04 | 0.025 | 0.025 |
| $Ecal isolation$ | 0.06 | 0.06 | 0.05 | 0.05 | 0.025 | 0.02 |
| $Hcal isolation$ | 0.05 | 0.03 | 0.025 | 0.025 | 0.02 | 0.02 |
| $Electron ID$ | | | | | | |
| $\sigma_{i\eta i\eta}$ | 0.03 | 0.03 | 0.03 | 0.03 | 0.03 | 0.03 |
| $\Delta\phi$ | 0.7 | 0.7 | 0.04 | 0.03 | 0.02 | 0.02 |
| $\Delta\eta$ | 0.01 | 0.009 | 0.007 | 0.007 | 0.005 | 0.005 |
| $H/E$ | 0.07 | 0.05 | 0.025 | 0.025 | 0.025 | 0.025 |

Table 3.1: Cut values at different working points in both barrel and endcap region

## 3.4 Background to $w \to e\nu$ events

Backgrounds [38] for electroweak boson production arise from two kinds of sources: isolated leptons originating from other electroweak boson production processes, and leptons (real or misidentified) originating from QCD jet or $\gamma + jet$ production.

**Electroweak background**

The electroweak background in the W sample consists mostly of $Z \to e^-e^+$ events with one electron escaping detection, and $W$ and $Z$ decays to $\tau's$ followed by a $\tau$ decay to an electron. Since these backgrounds are small, and because they arise from reliably computable electroweak cross sections, they can be estimated with adequate precision from simulation.

**Hadronic background**

The largest contribution to the background is expected to be from dijet events, in which one jet is misidentified as an electron while missing transverse energy results from mismeasurements. The size of the di-jet background depends on the probability of a jet faking an electron. This is hard to estimate and control from simulation, and therefore the QCD background must be measured from data. This background needs to be controlled carefully, as the uncertainty associated with it is larger than for the electroweak backgrounds mentioned above. Another background that contributes is $\gamma + jet$, where the electron candidate arises from a photon conversion in the detector.

### 3.4.1 Background study

To identify the electron we should know about the background sources which we can face while identifying the electrons. From the plot in Figure 3.2 we see that for MET<20 cut we get the maximum background and for MET>30 we get maximum signal. Plotting $E/p$ vs $f_{brem}$ helps us in understanding different type of background. We plot $E/p$ vs $f_{brem}$ in both barrel and endcap region, in endcap region the entries are more spreaded as compared to barrel region as shown by the histograms in Figure 3.3 and Figure 3.4, because the detector response is uniform in the barrel region. So dividing $E/p$ vs $f_{brem}$ plot into three regions(as shown in Figure 3.5) depending upon the histograms of $E/P$ and $f_{brem}$ which we can see in the next chapter. The cuts used for the divison are all hypothetical and will be discussed in the next chapter. After dividing into three region we will check the behavior of the electron identification variables in these three regions.

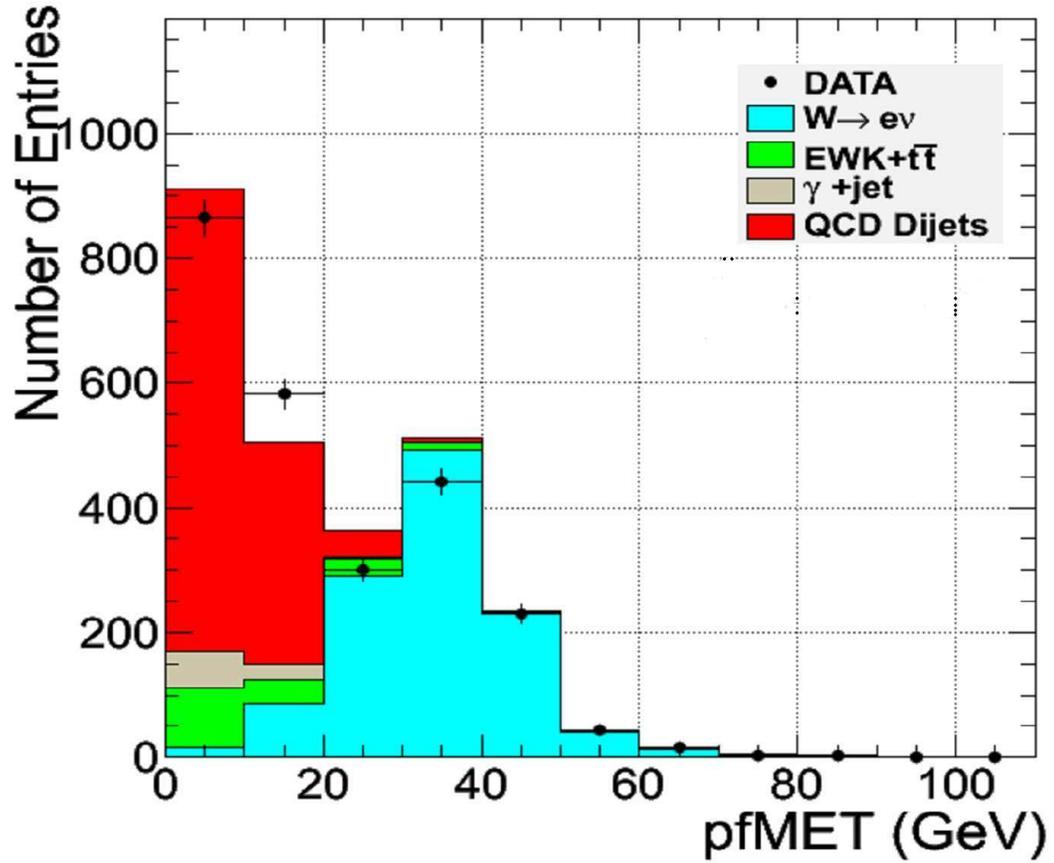

Figure 3-2: Histogram of Particle Flow Missing Transverse Energy (pfMET)

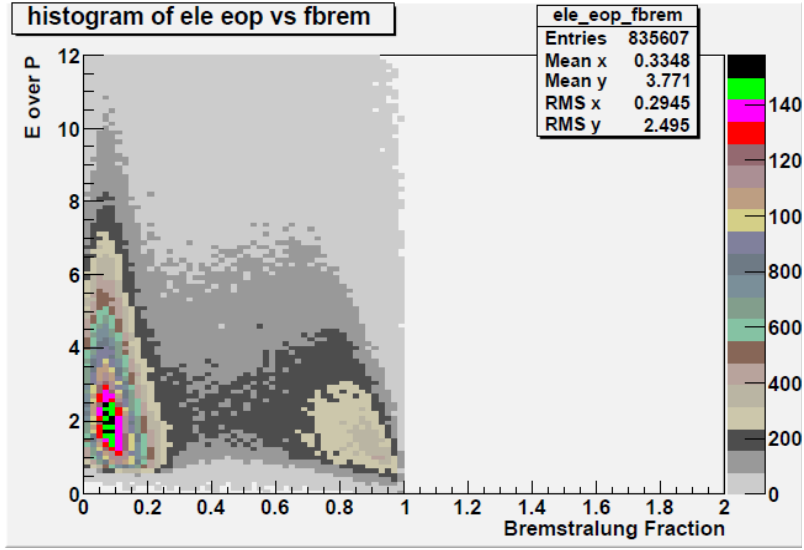

Figure 3-3: Histogram of $E/P$ vs $f_{brem}$ in endcap region for MET $< 20$.

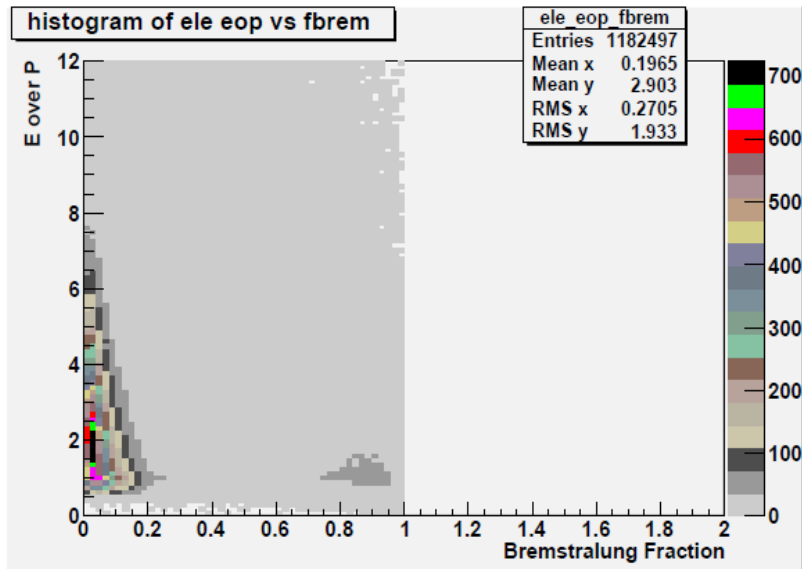

Figure 3-4: Histogram of $E/P$ vs $f_{brem}$ in barrel region for MET $< 20$

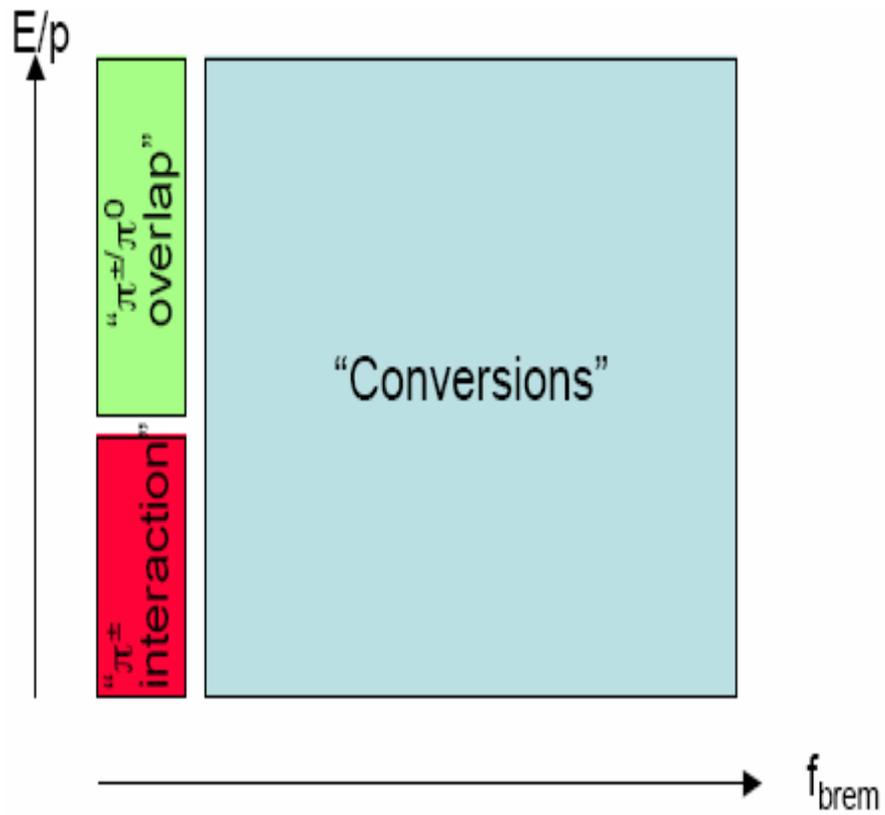

Figure 3-5: Divison of $E/P$ vs $f_{brem}$ plane in three regions

# Chapter 4

# Results and Discussion

## 4.1 Plots of electron identification variables without any selection

The electron identification variables are plotted by dividing the detector into two regions. The barrel region and the endcap region.

### 4.1.1 Barrel Region

The plots of electron identification variables in the barrel region without any selection are shown below.

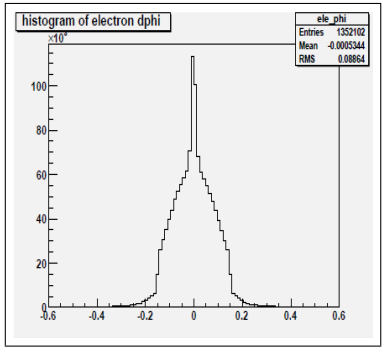 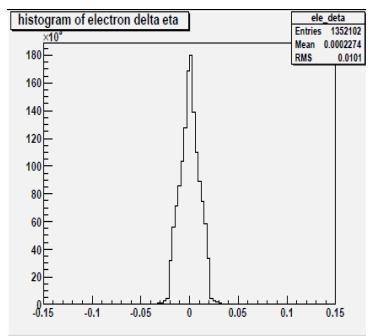 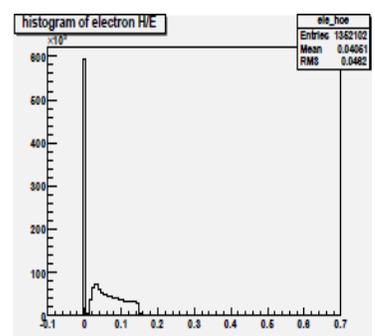

Figure 4-1: Plot of Delta phi(without any cut)

Figure 4-2: Plot of Delta eta(without any cut)

Figure 4-3: Plot of H/E(without any cut)

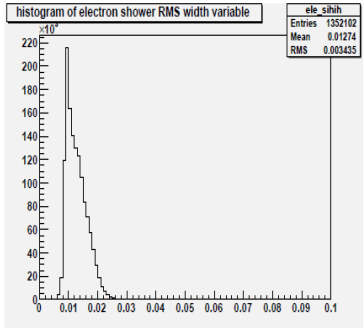 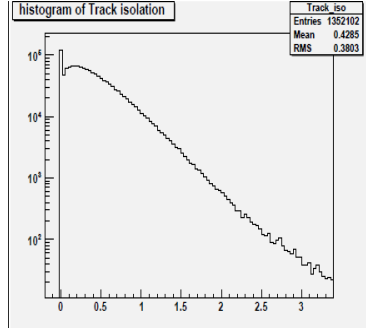 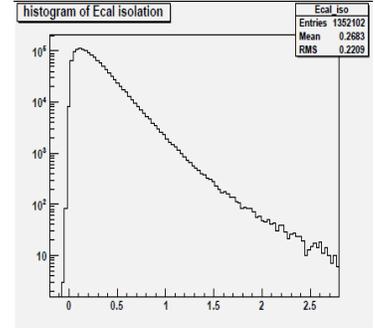

Figure 4-4: Plot of $\sigma_{i\eta i\eta}$(without any cut)

Figure 4-5: Plot of Track isolation(without any cut)

Figure 4-6: Plot of Ecal isolation(without any cut)

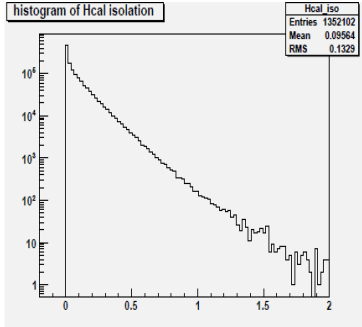 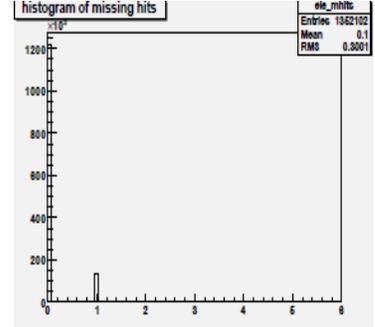

Figure 4-7: Plot of Hcal isolation(without any cut)

Figure 4-8: Plot of missing hits(without any cut)

### 4.1.2  Endcap Region

The plots of electron identification variables in the endcap region without any selection are shown below.

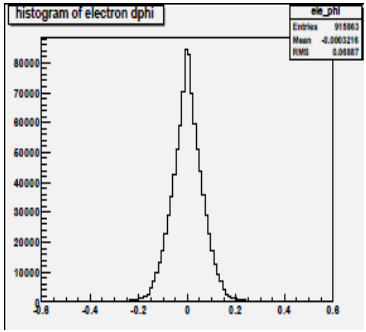

Figure 4-9: Plot of Delta phi(without any cut)

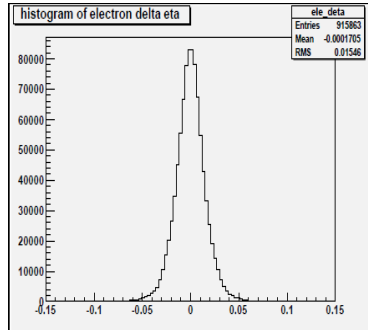

Figure 4-10: Plot of Delta eta(without any cut)

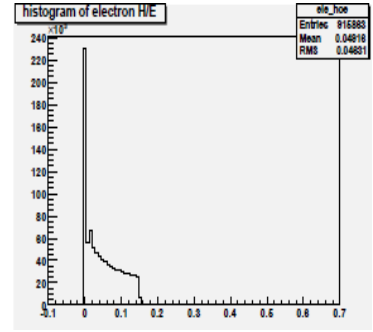

Figure 4-11: Plot of H/E(without any cut)

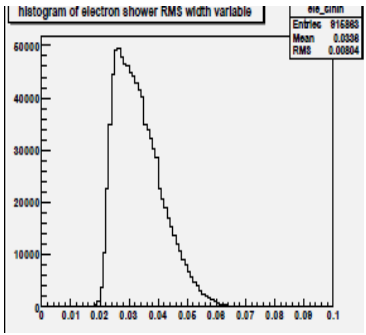

Figure 4-12: Plot of $\sigma_{i\eta i\eta}$(without any cut)

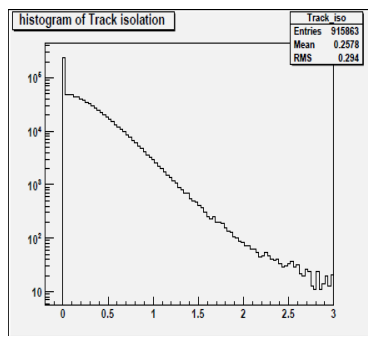

Figure 4-13: Plot of Track isolation(without any cut)

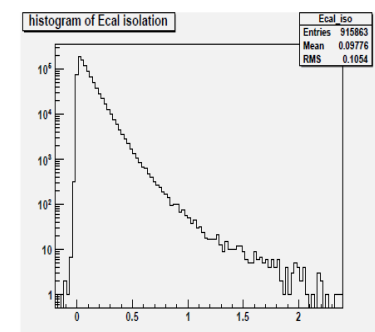

Figure 4-14: Plot of Ecal isolation(without any cut)

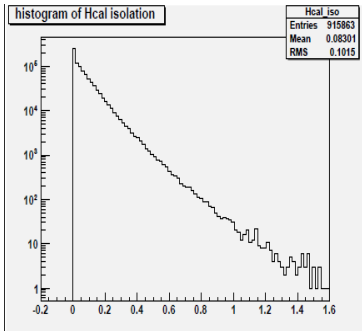

Figure 4-15: Plot of Hcal isolation(without any cut)

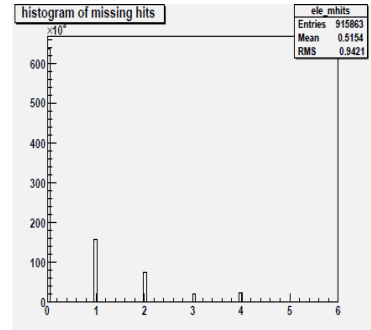

Figure 4-16: Plot of missing hits(without any cut)

## 4.2 N-1 plots of electron identification variables at working point 95(WP95)

Now we will see the N-1 plots of electron identification variables, N-1 plot means applying cuts on all the variables except the variable which is being plotted. The values of these cuts are given the Table 3.1.

### 4.2.1 Barrel Region.

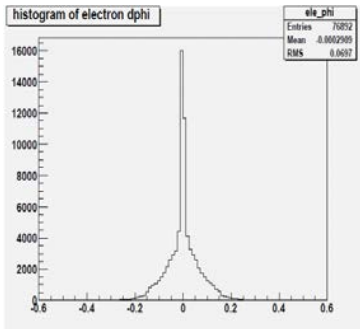

Figure 4-17: N-1 Plot of Delta phi(WP95)

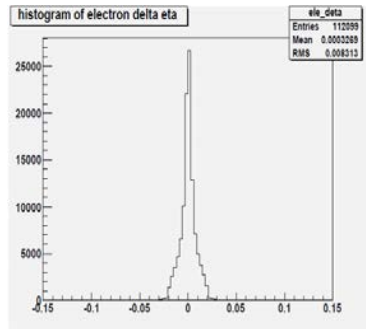

Figure 4-18: N-1 Plot of Delta eta(WP95)

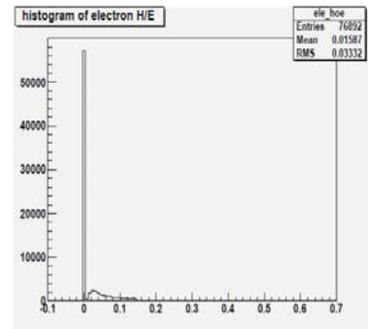

Figure 4-19: N-1 Plot of H/E(WP95)

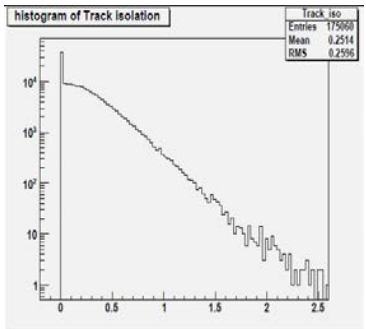

Figure 4-20: N-1 Plot of Track isolation(WP95)

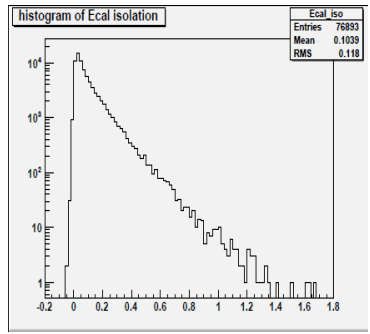

Figure 4-21: N-1 Plot of Ecal isolation(WP95)

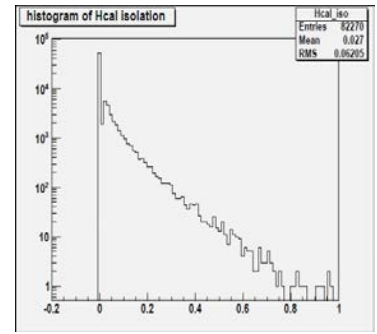

Figure 4-22: N-1 Plot of Hcal isolation(WP95)

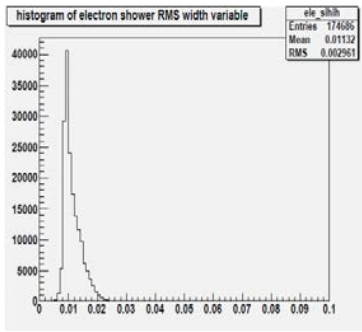

Figure 4-23: N-1 Plot of

$\sigma_{i\eta i\eta}$(WP95)

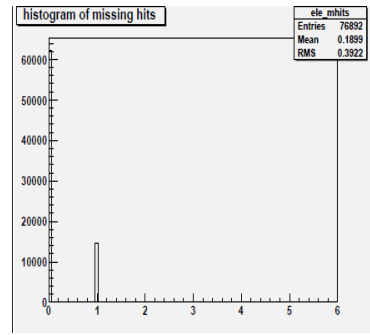

Figure 4-24: N-1 Plot of

missing hits(WP95)

### 4.2.2 Endcap Region

The plots of electron identification variables in the endcap region at WP95 are shown below

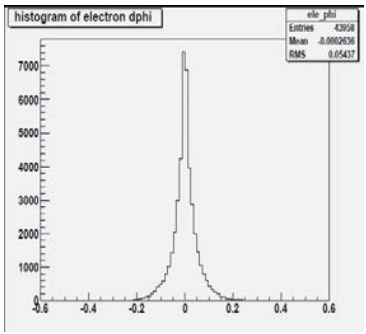

Figure 4-25: N-1 Plot of

Delta phi(WP95)

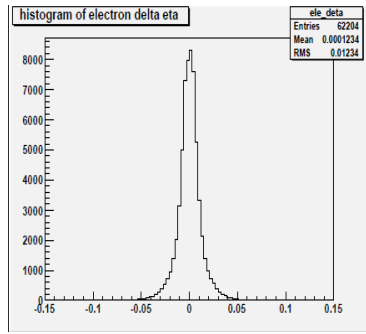

Figure 4-26: N-1 Plot of

Delta eta(WP95)

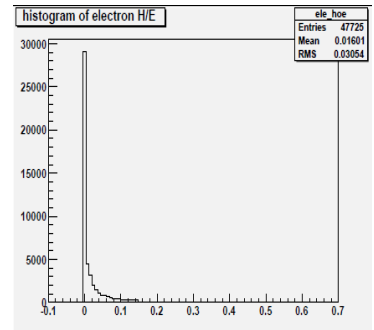

Figure 4-27: N-1 Plot of

H/EWP95)

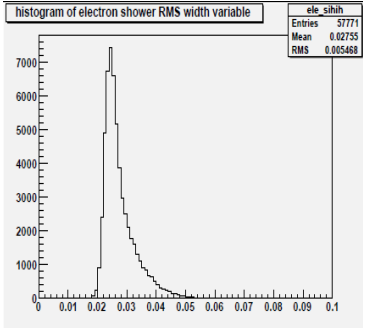
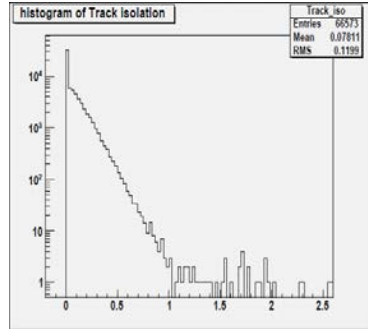
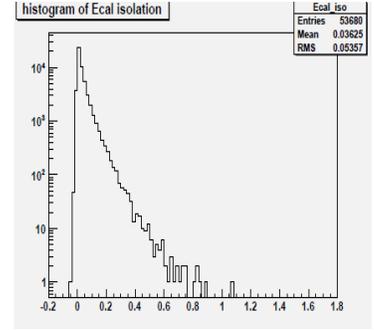

Figure 4-28: N-1 Plot of $\sigma_{i\eta i\eta}$(WP95)

Figure 4-29: N-1 Plot of Track isolation(WP95)

Figure 4-30: N-1 Plot of Ecal isolation(WP95)

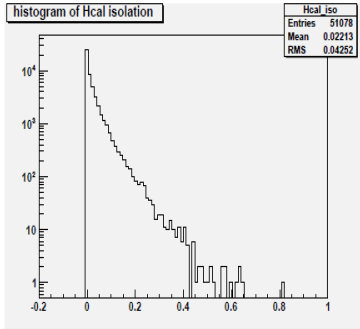
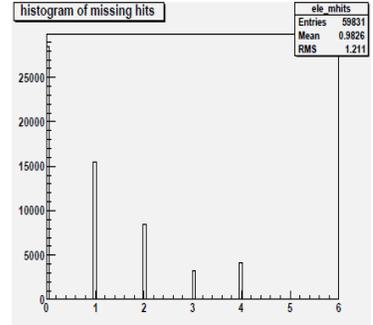

Figure 4-31: N-1 Plot of Hcal isolation(WP95)

Figure 4-32: N-1 Plot of missing hits(WP95)

## 4.3 N-1 plots of electron identification variables at working point 80(WP80)

Below are the N-1 plots of electron identification variables at working point 80 in both barrel and end cap region

### 4.3.1 Barrel Region.

The plots of electron identification variables in the barrel region at WP80 are shown below.

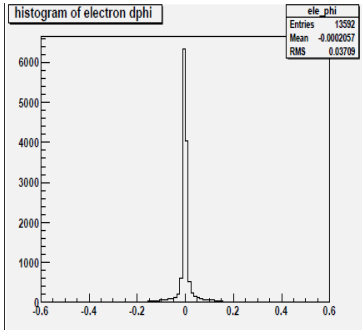

Figure 4-33: N-1 Plot of Delta phi(WP80)

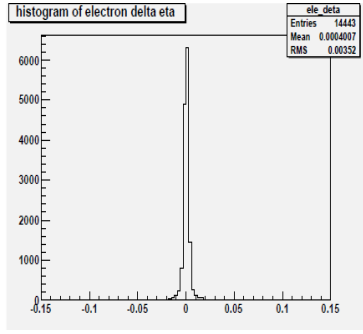

Figure 4-34: N-1 Plot of Delta eta(WP80)

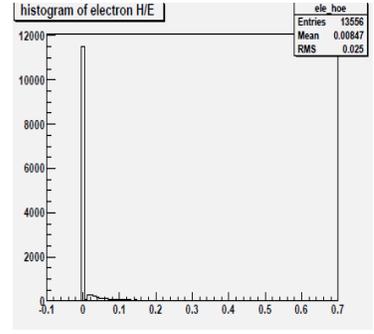

Figure 4-35: N-1 Plot of H/E(WP80)

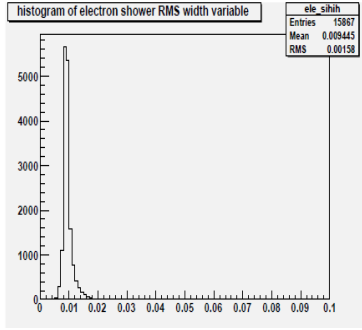

Figure 4-36: N-1 Plot of $\sigma_{i\eta i\eta}$(WP80)

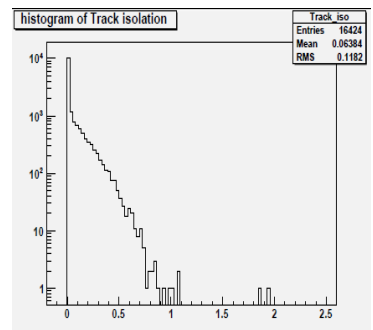

Figure 4-37: N-1 Plot of Track isolation(WP80)

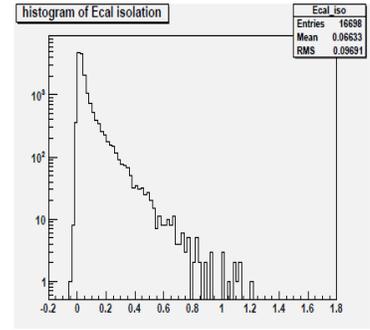

Figure 4-38: N-1 Plot of Ecal isolation(WP80)

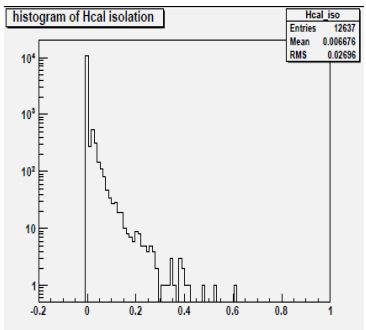

Figure 4-39: N-1 Plot of Hcal isolation(WP80)

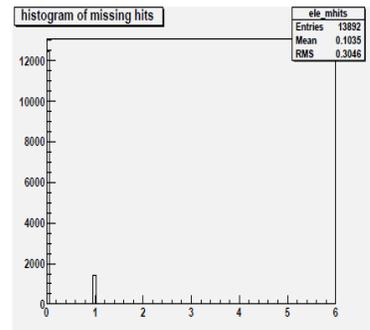

Figure 4-40: N-1 Plot of missing hits(WP80)

## 4.3.2 Endcap Region

The plots of electron identification variables in the endcap region at WP80 are shown below

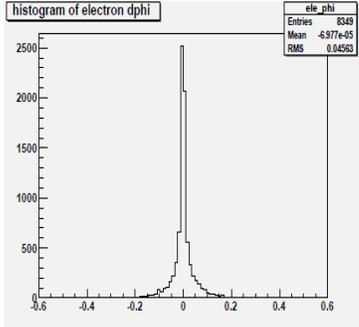

Figure 4-41: N-1 Plot of Delta phi(WP80)

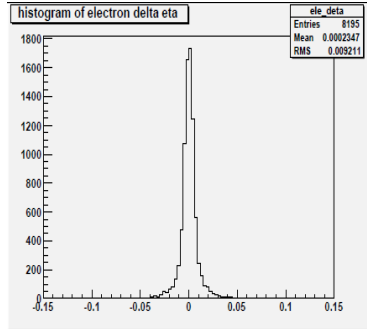

Figure 4-42: N-1 Plot of Delta eta(WP80)

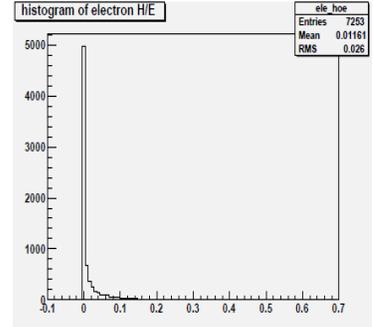

Figure 4-43: N-1 Plot of H/E(WP80)

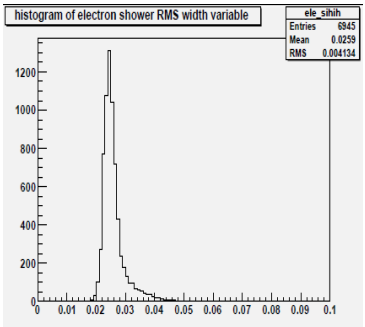

Figure 4-44: N-1 Plot of $\sigma_{i\eta i\eta}$(WP80)

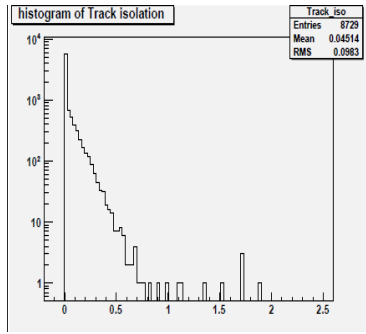

Figure 4-45: N-1 Plot of Track isolation(WP80)

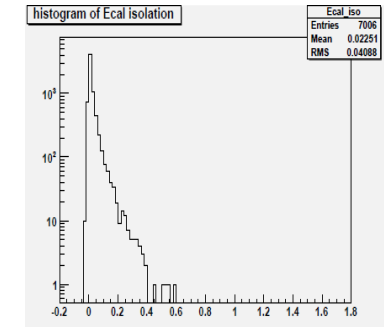

Figure 4-46: N-1 Plot of Ecal isolation(WP80)

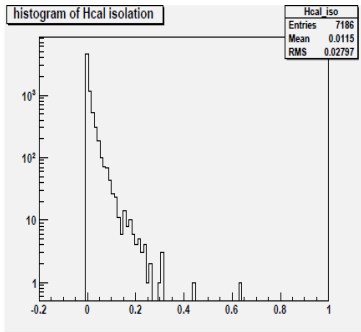
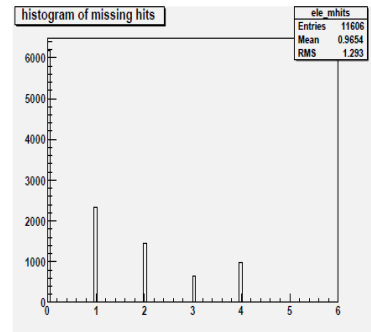

Figure 4-47: N-1 Plot of Hcal isolation(WP80)

Figure 4-48: N-1 Plot of missing hits(WP80)

## 4.4 N-1 plots of electron identification variables at working point 60(WP60)

Below are the N-1 plots of electron identification variables at working point 60 in both barrel and end cap region

### 4.4.1 Barrel Region

The plots of electron identification variables in the barrel region at WP60 are shown below.

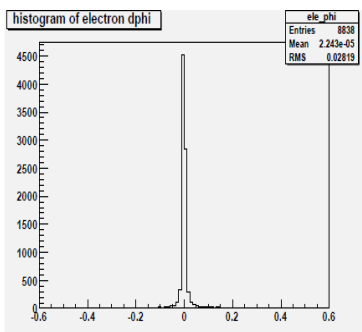
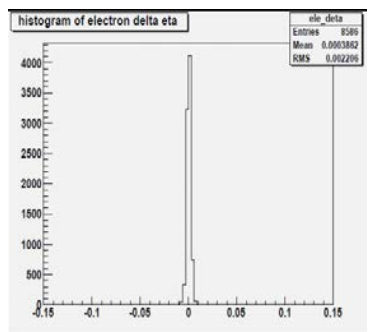
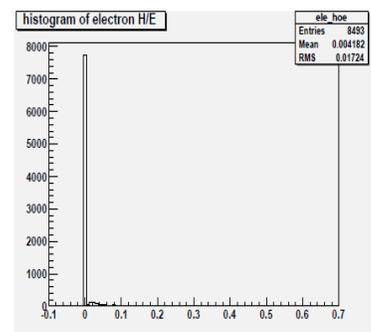

Figure 4-49: N-1 Plot of Delta phi(WP60)

Figure 4-50: N-1 Plot of Delta eta(WP60)

Figure 4-51: N-1 Plot of H/E(WP60)

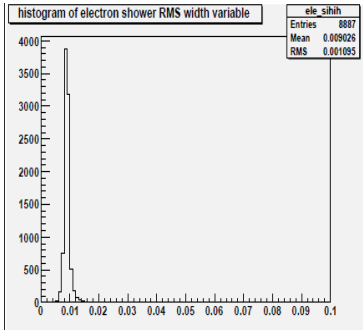

Figure 4-52: N-1 Plot of

$\sigma_{i\eta i\eta}$(WP60)

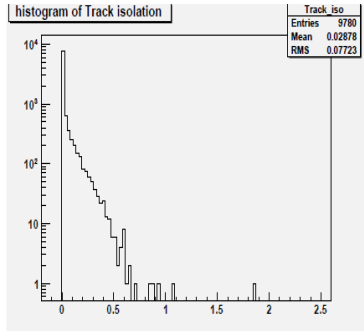

Figure 4-53: N-1 Plot of

Track isolation(WP60)

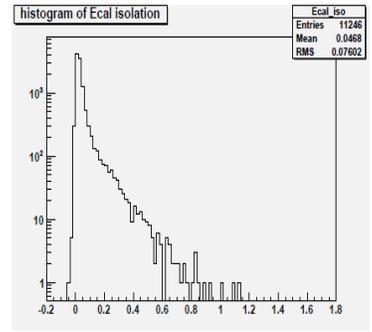

Figure 4-54: N-1 Plot of

Ecal isolation(WP60)

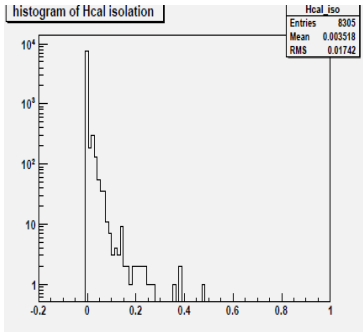

Figure 4-55: N-1 Plot of

Hcal isolation(WP60)

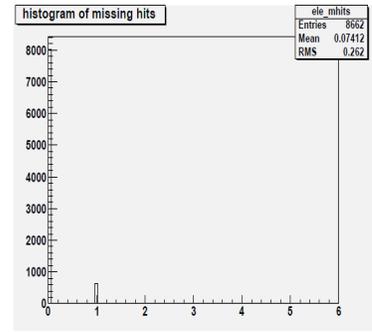

Figure 4-56: N-1 Plot of

missing hits(WP60)

### 4.4.2 Endcap Region

The plots of electron identification variables in the endcap region at WP60 are shown below

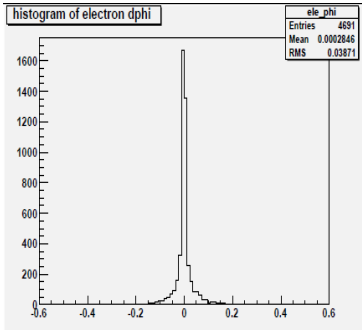

Figure 4-57: N-1 Plot of Delta phi(WP60)

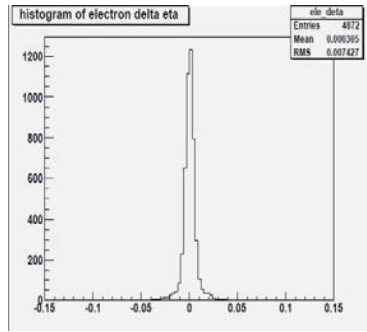

Figure 4-58: N-1 Plot of Delta eta(WP60)

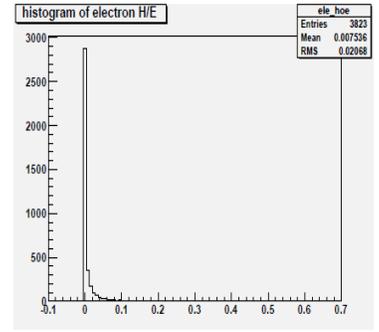

Figure 4-59: N-1 Plot of H/E(WP60)

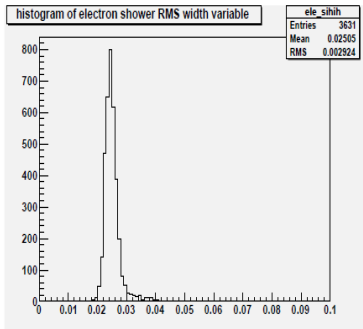

Figure 4-60: N-1 Plot of $\sigma_{i\eta i\eta}$(WP60)

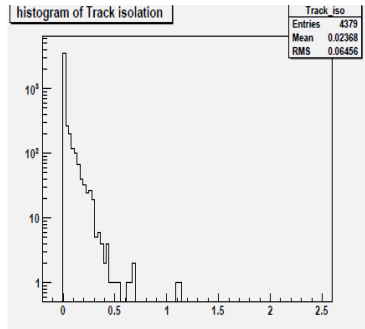

Figure 4-61: N-1 Plot of Track isolation(WP60)

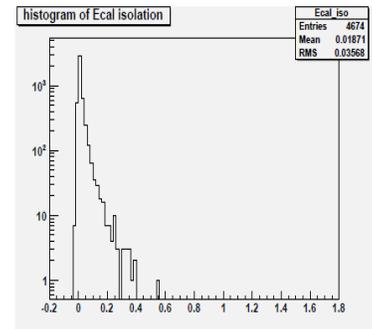

Figure 4-62: N-1 Plot of Ecal isolation(WP60)

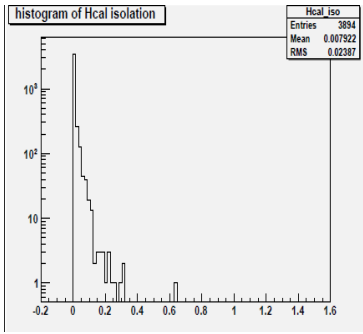
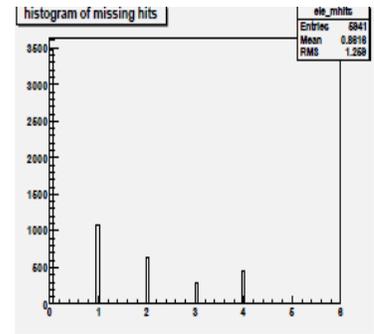

Figure 4-63: N-1 Plot of Hcal isolation(WP60)

Figure 4-64: N-1 Plot of missing hits(WP60)

## 4.5 Categorization of $E/P$ vs $f_{brem}$ plot for the background study

$E/P$ vs $f_{brem}$ plot is divided into three regions to study the background. Dividing into categories helps us in understanding different sources of background. Below are the plots of $f_{brem}$ and $E/P$ which will help us in dividing $E/P$ vs $f_{brem}$ plot into three regions.

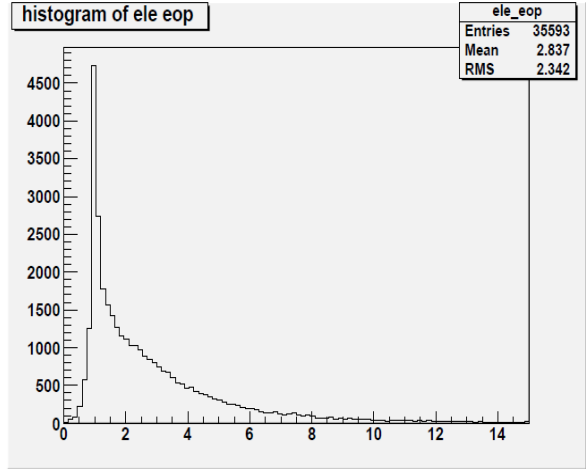

Figure 4-65: Plot of E/P in the barrel region for signal(MET > 30)

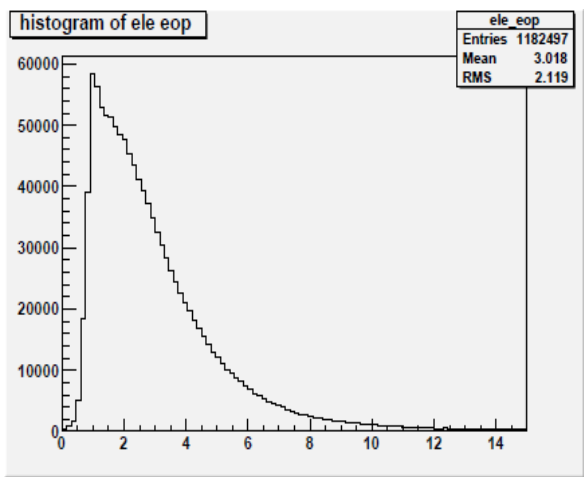

Figure 4-66: Plot of E/P in the barrel region

for background(MET < 20)

In both Figures 4-65 and 4-66 we get a peak at about $E/P \cong 1$, for signal this peak is little narrow and for background the peak is wider. So we can divide E/P into two region, $E/P > 1$ and $E/P < 1$. Now we will see the plot of $f_{brem}$ for both signal and background in the barrel region.

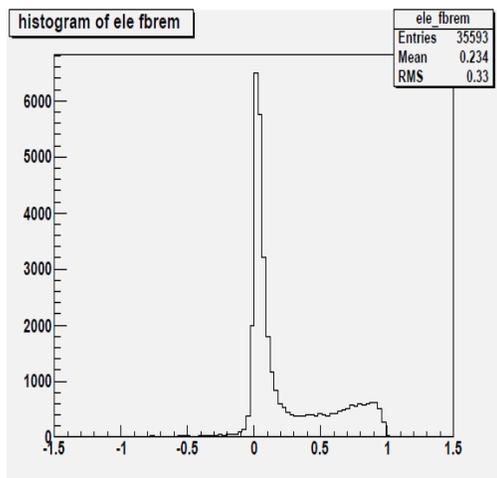

Figure 4-67: Plot of $f_{brem}$ in the barrel region for signal(MET > 30)

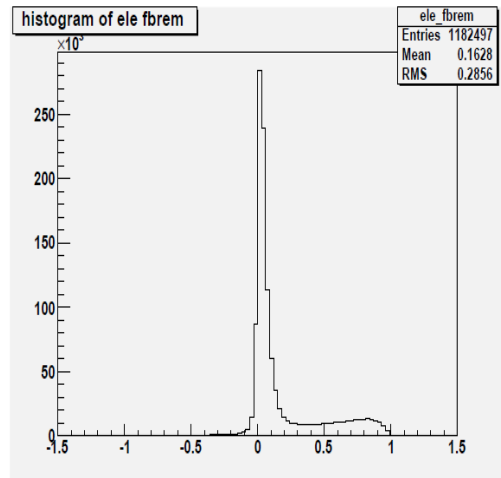

Figure 4-68: Plot of $f_{brem}$ in the barrel region for background(MET < 20)

In both Figures 4-67 and 4-68 we get a peak at $f_{brem} \cong 0.15$. So $f_{brem}$ plot can be divided into two regions i-e $f_{brem}$ >0.15 and $f_{brem}$ <0.15. Below are the plots of E/P in the barrel region for background but with $f_{brem}$ cut applied.

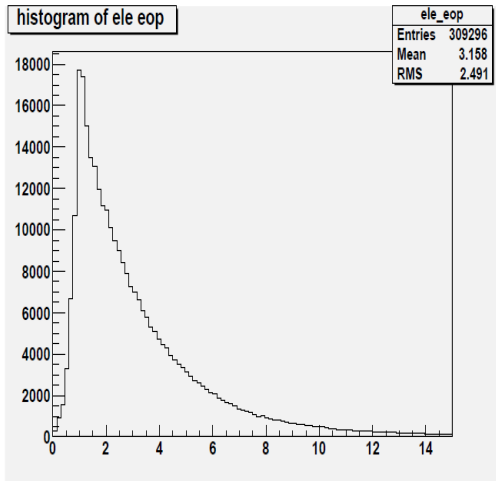 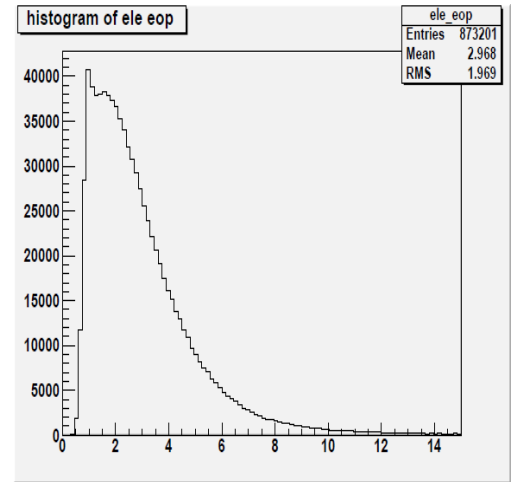

Figure 4-69: Plot of $E/P$ in the barrel region for background(MET < 20) and $f_{brem}$ >0.15

Figure 4-70: Plot of $E/P$ in the barrel region for background(MET < 20) and $f_{brem}$ <0.15

After having a look at the plots of $E/P$ and $f_{brem}$ we can divide the $E/P$ vs $f_{brem}$ plot into three regions. Below are the cuts used for divison.

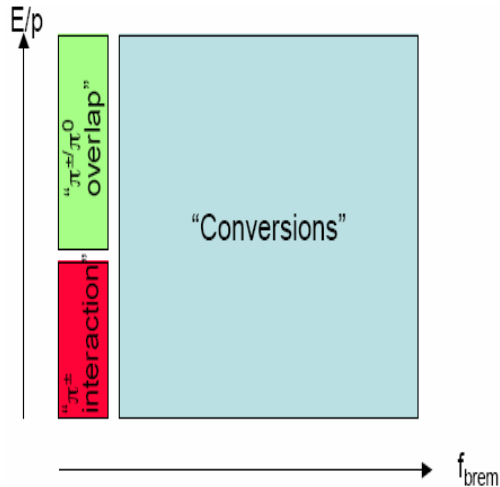

Figure 4-71: Divison of $E/P$ vs $f_{brem}$ plane in three regions

- Region A $\implies f_{brem} > 0.15$.
- Region B $\implies f_{brem} <0.15$ & $E/P < 1$.
- Region C $\implies f_{brem} <0.15$ & $E/P > 1$.

All of these cuts (cuts for divison into three regions) are chosen hypothetically after analyzing the $E/P$ and $f_{brem}$ distributions. There is also a hypothesis about the sources of background in these three regions.

- In region A the background is mainly due to conversion of photons into electrons.
- In region B charged hadron probably $\pi^{\pm}$ showers early in the tracker and so transmitt a small amount of energy in Ecal and thus have low $E/P$.
- In region C, $E/P > 1$, the reason is because the energy deposited by $\pi^0$ in Ecal overlap accidently with the track of $\pi^{\pm}$.

## 4.6 Behavior of Electron Identification variables in Region A,B and C

Now we want to study how the electron identification variables behaves in these three regions separately. So we will plot $\Delta\phi$, $\Delta\eta$ and $\sigma_{i\eta i\eta}$ in the three regions.

### 4.6.1 Plot of $\Delta\phi$ in the three regions without any selection

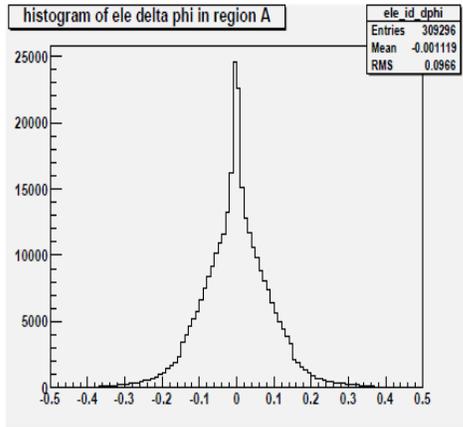
Figure 4-72: $\Delta\phi$ in Region A(without any selection)

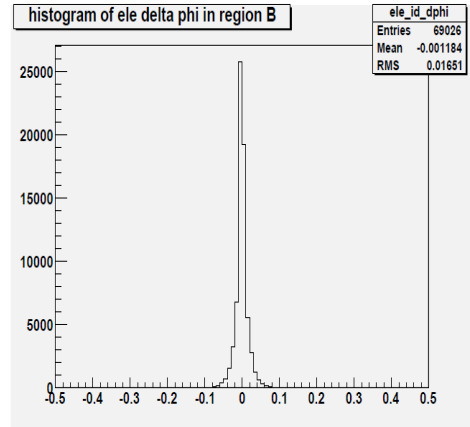
Figure 4-73: $\Delta\phi$ in Region B(without any selection)

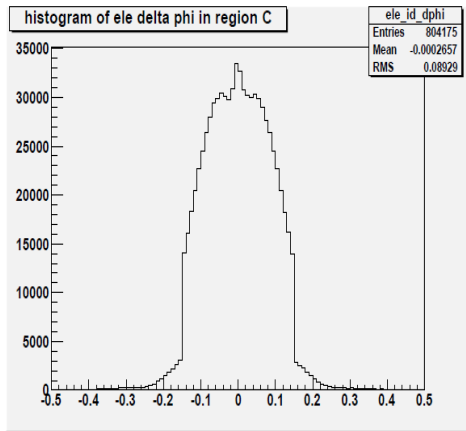

Figure 4-74: $\Delta\phi$ in Region C(without any selection)

### 4.6.2 Plot of $\Delta\eta$ in the three regions without any selection

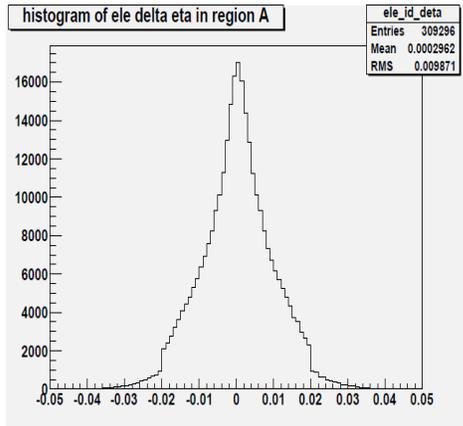 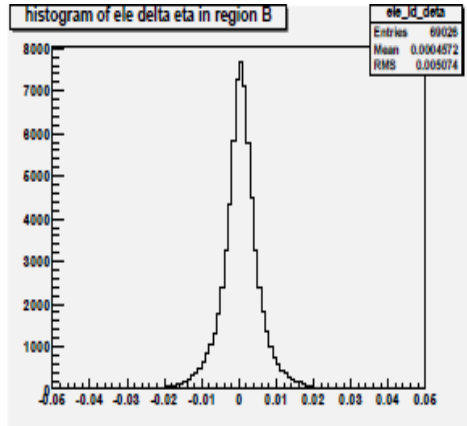

Figure 4-75: $\Delta\eta$ in Region A(without any selection)

Figure 4-76: $\Delta\eta$ in Region B(without any selection)

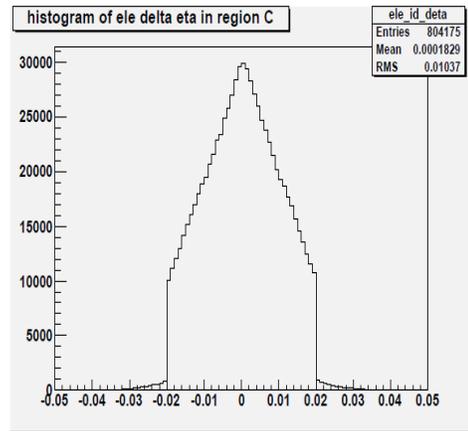

Figure 4-77: $\Delta\eta$ in Region C(without any selection)

### 4.6.3 Plot of $\sigma_{i\eta i\eta}$ in the three regions without any selection

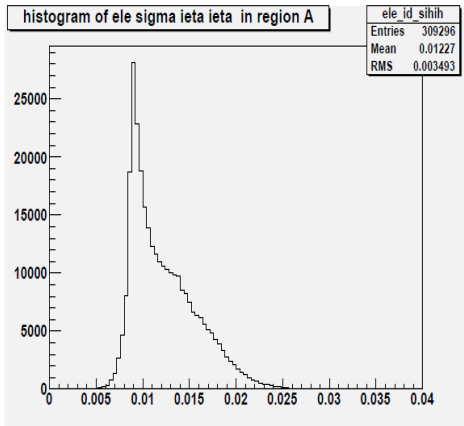

Figure 4-78: $\sigma_{i\eta i\eta}$ in Region A(without any selection)

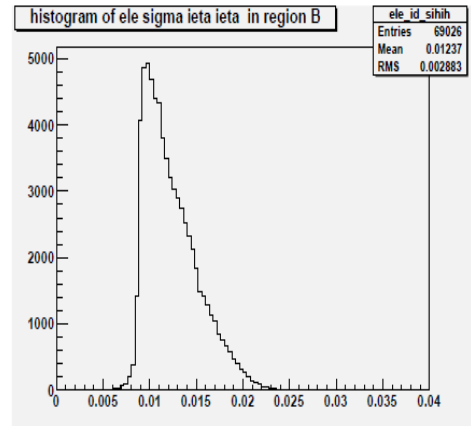

Figure 4-79: $\sigma_{i\eta i\eta}$ in Region B(without any selection)

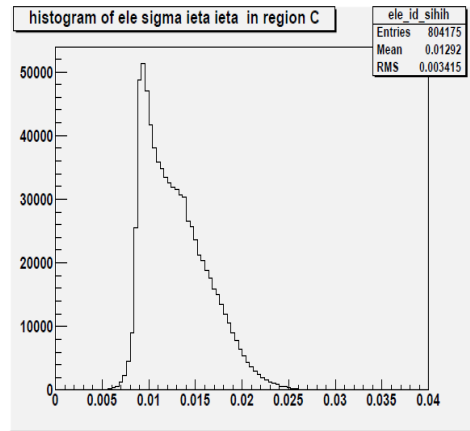

Figure 4-80: $\sigma_{i\eta i\eta}$ in Region C(without any selection)

Now we will see N-1 plots (with WP80 cuts) of theses varaibles in the three regions.

### 4.6.4 N-1 Plot of $\Delta\phi$ in the three regions

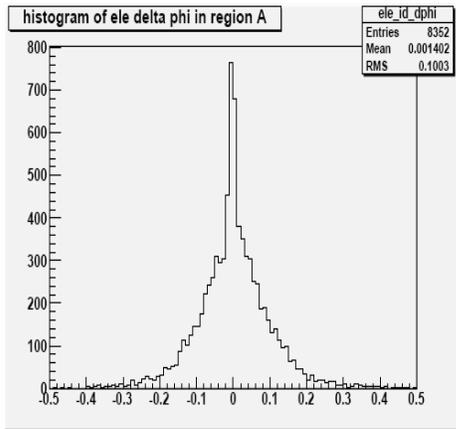
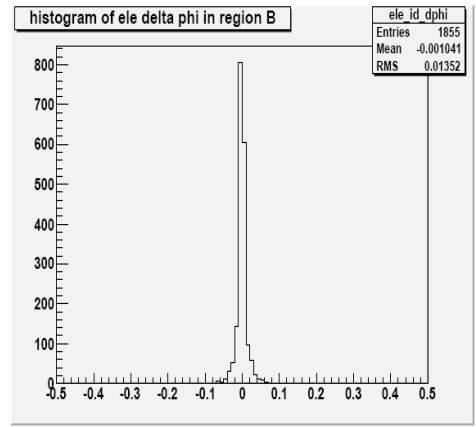

Figure 4-81: N-1 plot of $\Delta\phi$ in Region A(WP80)

Figure 4-82: N-1 plot of $\Delta\phi$ in Region B(WP80)

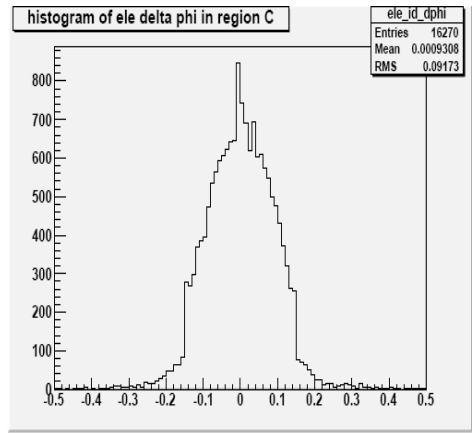

Figure 4-83: N-1 plot of $\Delta\phi$ in

Region C(WP80)

## 4.6.5 N-1 Plot of $\Delta\eta$ in the three regions

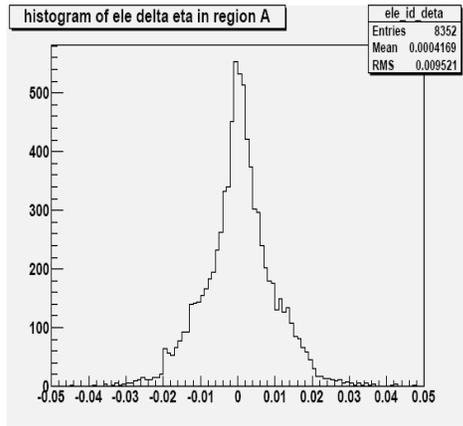 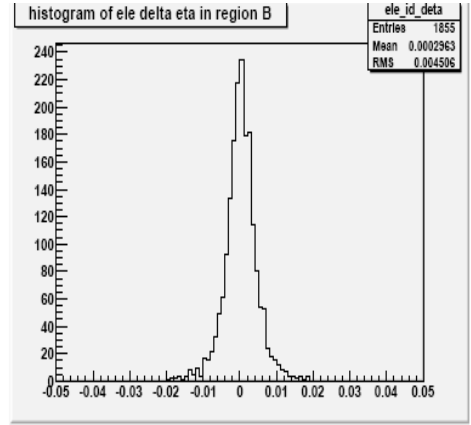

Figure 4-84: N-1 plot of $\Delta\eta$ in

Region A(WP80)

Figure 4-85: N-1 plot of $\Delta\eta$ in

Region B(WP80)

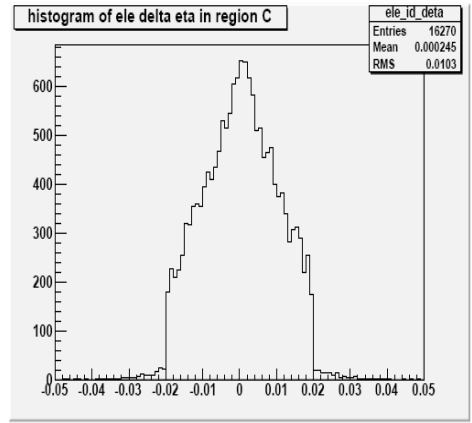

Figure 4-86: N-1 plot of $\Delta\eta$ in

Region C(WP80)

### 4.6.6 N-1 Plot of $\sigma_{i\eta i\eta}$ in the three regions

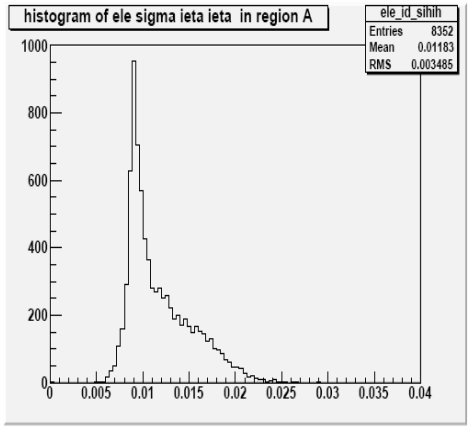

Figure 4-87: N-1 plot of $\sigma_{i\eta i\eta}$ in

Region A(WP80)

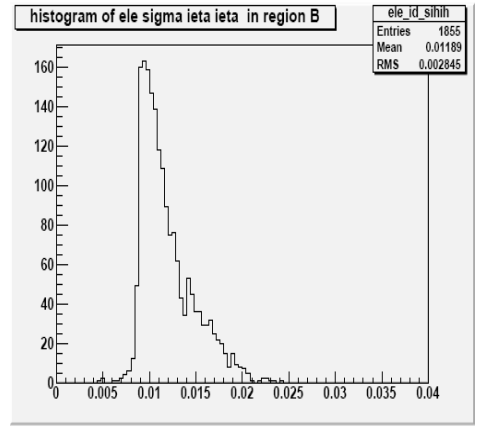

Figure 4-88: N-1 plot of $\sigma_{i\eta i\eta}$ in

Region B(WP80)

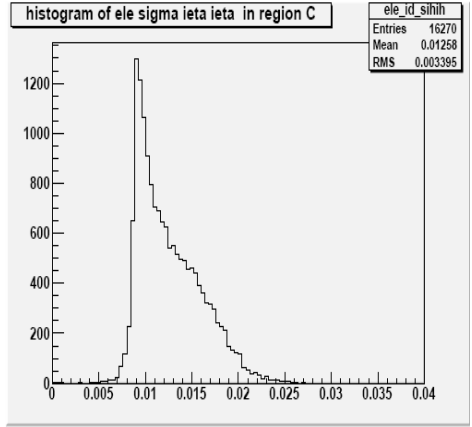

Figure 4-89: N-1 plot of $\sigma_{i\eta i\eta}$ in

Region C(WP80)

## 4.7 Data set Used

| Dataset | Runs | luminosity | JSON in use |
|---|---|---|---|
| /MinimumBias/Commissioning10 | 131511- | $8.0\text{nb}^{-1}$ | 132440-137028_7TeV_June |
| -SD_EG-Jun14thSkim_v1/RECO | 135802 | | 14thReReco_Collisions10_JSON_v2 |
| /EG/Run2010A-Jun14thReReco | 135821- | $4.9\text{nb}^{-1}$ | 132440-137028_7TeV_June14thRe |
| _v1/RECO | 137436 | | Reco_Collisions10_JSON_v2 |
| /EG/Run2010A-PromptReco | 137437- | $61.2\text{nb}^{-1}$ | N132440-141961_7TeV_Stream |
| -v4/RECO | 139558 | | Express_Collisions10_JSON |
| | 140160- | $5.2\text{nb}^{-1}$ | 132440-141961_7TeV_Stream |
| | 140182 | | Express_Collisions10_JSON |
| /EG/Run2010A-Jul16thReReco | 139559- | $118.8\text{nb}^{-1}$ | 139779-1340159_7TeV_July16thRe |
| -v2/RECO | 140174 | | Reco_Collisions10_JSON |
| /EG/Run2010A-PromptReco | 140183- | $1.13\text{pb}^{-1}$ | 132440-143336_7TeV_Stream |
| -v4/RECO | 143336 | | Express_Collisions10_JSON_v2 |
| /EG/Run2010A-PromptReco | 143337- | $1.54\text{pb}^{-1}$ | 132440-141961_7TeV_Stream |
| -v4/RECO | 144144 | | Express_Collisions10_JSON |
| **Total** | | $\mathbf{2.8\text{pb}^{-1}}$ | |

Table 4.1: Dataset used for the analysis

## 4.8 Conclusion

Cut based method is used to identify the electrons and to minimize the background with a very small loss of signal. Large and varying amount of tracker makes electron identification a difficult task. Electron identification variables are plotted seperately in barrel and endcap regions at different working points. Each working point introduces different set of cuts for barrel and endcap. In the endcap region the cuts are little tight as compared to the barrel region as the background is larger in endcap and also the detector response is not uniform. $\Delta\phi$ and $\Delta\eta$ should be approximately zero for real electron, because the track and cluster energy should match with each other in both $\phi$ and $\eta$. H/E should also be nearly equal to zero for electrons as they deposit almost all of their energy in the Ecal, $\sigma_{i\eta i\eta}$ should be 0.01 in the barrel and 0.03 in the endcap because in the endcap we have bigger crystals as compared to the barrel. Track isolation, Ecal isolation and Hcal isolation variables must be close to zero inorder to isolate the electrons. Missing hits for electrons should also be less. The background is reduced as we use more tight cuts and the signal becomes more prominent. WP60 introduce more tight set of cuts than WP80 and similarly WP80 set of cuts are more tighter than WP95. In this way such variables are introduced which can give us over desired electrons i-e electrons coming from W boson.

An attempt to study the background has been made to distinguish real electrons from the fake electrons. This method has been used for the first time on LHC real data to study the background sources for electrons. The distribution of $\Delta\phi$ in region C is very wide as shown in Figures 4-74 and 4-83, because the track and shower overlap accidently. In region B there is a good track cluster match as shown by the plots in Figure 4-73 and Figure 4-82, because the shower is caused by the charged hadrons, whereas $\phi$ measurement is spoilt by radiation in tracker in region A as shown in Figure 4-72 and Figure 4-81. Same is the situation with the distributions of $\Delta\eta$ in region A,B and C, track cluster match in $\eta$ is good in region B as compared to region A and C. The shower width variable $\sigma_{i\eta i\eta}$ behaves differently in the three regions. In region A we get a peak at about $\sigma_{i\eta i\eta} \approx 0.01$ showing the clean conversion electrons as shown in Figure 4-78 and Figure 4-87. As the hadronic showers are little wide thats why we get a broader distribution of $\sigma_{i\eta i\eta}$ in the region B as shown by the plots in Figure 4-79 and Figure 4-88. In region C the peak in the distribution of $\sigma_{i\eta i\eta}$ shows overlap of single $\pi^0$ with $\pi^\pm$ and the wider portion shows the overlap of multi $\pi^0$ with $\pi^\pm$ as shown by the plots in Figure 4-80 and Figure 4-89. After having a look at the plots one can say that region B is a region of real electrons but its not possible because we are applying MET $< 20$ cut which means we are working in background so it implies that region B is a source of good fake electrons.